\documentclass[%
 reprint,
 showpacs,preprintnumbers,
 amsmath,amssymb,
 aps,
 prb,
]{revtex4-1}
\usepackage{dcolumn}
\usepackage[dvipdfm]{graphicx}
\usepackage{subfigure,color}
\usepackage{mathrsfs}

\makeatletter
\def\btt#1{\texttt{\@backslashchar#1}}
\DeclareRobustCommand\bblash{\btt{\@backslashchar}} \makeatother

\begin{document}

\title{Robustness of Gapless Interface State in a Junction of Two Topological Insulators}

\author{Tetsuro Habe$^1$ and Yasuhiro Asano$^{1,2}$}
\affiliation{$^1$Department of Applied Physics,
Hokkaido University, Sapporo 060-8628, Japan}
\affiliation{$^2$Center for Topological Science \& Technology,
Hokkaido University, Sapporo 060-8628, Japan}

\date{\today}

\begin{abstract}
We theoretically study 
subgap states appearing at the interface between two three-dimensional topological insulators which have different 
configurations in the spin-orbit 
interactions from each other. The coupling of spin $\boldsymbol{\sigma}$ with momenta $\boldsymbol{p}$ 
is configured by a material dependent $3\times 3$ matrix $\boldsymbol{\Lambda}$ as ${\sigma}^\mu {\Lambda}_\mu^\nu p_\nu$.
The spectra of the interface subgap states depend on the relative choices of $\boldsymbol{\Lambda}$'s in the two topological insulators where the two $\boldsymbol{\Lambda}$ are connected by the unitary transformation including the inversion and the rotation in momentum space.
The gapless states appear at the interface when the transformation includes the inversion.
The two topological insulators can be distinguished by using the topological numbers
defined in the two- or one-dimensional partial Brillouin zone, which explains the presence of the gapless
interface states.
We also discuss the robustness of such gapless states under perturbations breaking the time-reversal symmetry or the band-inversion symmetry.
\end{abstract}

\pacs{73.20.At, 74.45.+c}

\maketitle

\section{introduction}
Topological insulators (TIs) are new class of condensed matter\cite{Fu2007,Fu2007-2,Moore2007,Chen2009}. 
A topological number $Z_2$ defined in terms of the global property of wave function for
the occupied states below the gap distinguishes the topological insulating phase $Z_2=1$
from the conventional one $Z_2=0$. 
The strong spin-orbit interaction locks the direction 
of momenta and that of spin, and is responsible for the non-trivial $Z_2$ topological number.
The bulk-edge correspondence guarantees the presence 
of gapless states at the surface of TIs\cite{Fu2007,Moore2007,Roushan2009,T.Zhang2009}.
On the surface of three-dimensional TIs, 
the excitation spectra of the gapless state are described by the so-called Dirac cone,
(i.e., $E=\pm v |\boldsymbol{p}|$). 

When we focus on the surface state of a single TI, the spectra of the surface state are independent of
the configurations in the momentum-spin locking. 
On the other hand,
when we focus on the interface states between two different topological materials,
the spectra of the interface state depend on the relative configuration 
of the spin-orbit coupling in the two TIs~\cite{Takahashi2011,Sen2012} and superconductors\cite{Barash2001,Kwon2004,Asano2010}.
These studies focus on the discrete degree of freedom such as the helicity, the chirality 
and the mirror symmetry~\cite{Teo2008,Nishide2010} to characterize the relative configuration of the spin-orbit coupling.
However, the spin-orbit coupling allows more complex relative configuration which is represented by a material dependent $3\times 3$ matrix $\boldsymbol{\Lambda}$ configuring the coupling of spin $\sigma^\mu$ with momenta $p_\nu$ as 
${\sigma}^\mu {\Lambda}_\mu^\nu p_\nu$. 
In junctions, generally speaking, $\boldsymbol{\Lambda}_{(1)}$ in one TI and $\boldsymbol{\Lambda}_{(2)}$ in the other are connected  by the transformation (i.e., $\boldsymbol{\Lambda}_{(1)}=\boldsymbol{\Lambda}_{12}\boldsymbol{\Lambda}_{(2)}$ ) not only by the discrete inversion in momentum space but also by 
the continuous rotation there.
The properties of the interface subgap states 
would depend on $\boldsymbol{\Lambda}_{12}$ which represents the relative choice of $\boldsymbol{\Lambda}$ in the two TIs. 

In this paper, we discuss the properties of the two-dimensional states 
appearing at the interface of two three-dimensional TIs 
characterized by the two different $\boldsymbol{\Lambda}$.
The matrix $\boldsymbol{\Lambda}_{12}$ includes the two transformations: 
(i) the inversion in momentum space 
and 
(ii)
the continuous rotation of the momentum-spin locking angle ($\phi$) . 
The stability of the gapless interface state depends on the structure of the transformation represented by $\boldsymbol{\Lambda}_{12}$.
The different topological numbers characterize the gapless interface states 
depending on whether or not the inversion includes the momentum perpendicular to the interface.

The inversion in momentum space is categorized into two cases in terms of the determinant of $\boldsymbol{\Lambda}_{12}$.
Namely $\mathrm{det}[\boldsymbol{\Lambda}_{12}]=-1(+1)$ distinguish the transformation including the inversion in odd number of momenta axes (even number of momenta axes).
When $\boldsymbol{\Lambda}_{12}$ includes the inversion in two momenta parallel to the junction plane ($\mathrm{det}[\boldsymbol{\Lambda}_{12}]=1$), 
the zero-energy interface states appear. But such zero-energy states are fragile under the rotation of momenta within the interface plane.
We conclude that the gapless interface states between two TIs are characterized by the relative Chern number, which has the robustness of the gapless interface states under the Zeeman field for all spin directions.
When $\boldsymbol{\Lambda}_{12}$ includes the inversion along the axis perpendicular to the junction plane, we find the gapless interface state characterized
by the Sato's winding number~\cite{SatoM2010,Yada2011,SatoM2011,Mizushima2012}.
It has been already known that the surface state of a single TI is sensitive to the direction of 
Zeeman field~\cite{Tanaka2009,Liu2009,Wray2010,Habe2012}.
We also find that the interface states at the zero energy are also sensitive to the direction of 
Zeeman field. We conclude that such magnetic anisotropy stems from the mirror symmetry.
Unfortunately, all of the gapless interface states are fragile under the perturbations 
which break the band-inversion symmetry of two TIs in a different way.

This paper is organized as follows. In Sec.~\ref{model}, we explain the theoretical model considered in this paper. 
In Sec.~III, we show the two topological numbers which guarantee the gapless interface state in the junction.
In Sec.~IV, we confirm the topological analysis and the robustness
of the zero-energy interface states by the numerical simulation on the tight-binding model.
The conclusion is given in Sec.~V.

\section{model}\label{model}
The most simple Hamiltonian of three-dimensional topological insulator is given by,
\begin{align}
H_{\textrm{TI}}=\begin{pmatrix}
(m-b\boldsymbol{p}^2)\sigma^0& a \boldsymbol{\sigma}\cdot\boldsymbol{p}\\
a \boldsymbol{\sigma}\cdot\boldsymbol{p}& -(m-b\boldsymbol{p}^2)\sigma^0
\end{pmatrix},
\end{align}
where $a$, $b$, and $m$ are positive constants, $\sigma^0$ is $2\times2$ identity matrix, and 
$\boldsymbol{\sigma}=(\sigma^x,\sigma^y,\sigma^z)$ are Pauli matrices in spin space.
The eigen values of the Hamiltonian are $\pm E_p$ with $E_p=\sqrt{M^2+a^2|\boldsymbol{p}|^2}$ and $M=(m-b\boldsymbol{p}^2)$.
For convenience, we utilize the short notation as,
\begin{align}
H_{\textrm{TI}}=a\alpha^{\mu}p_{\mu}+M\beta,\;\;\;\mu=x,y,z, \label{Hamiltonian}
\end{align}
where $M$ is regarded as the Dirac mass and we use $4\times4$ Dirac matrices,
\begin{align*}
\alpha^{\mu}=\begin{pmatrix}
0&\sigma^{\mu}\\
\sigma^{\mu}&0
\end{pmatrix},\;\;\;
\beta=\begin{pmatrix}
\sigma^0&0\\
0&-\sigma^0
\end{pmatrix}.
\end{align*}
Eq.~(\ref{Hamiltonian}) is called Dirac Hamiltonian and describes electronic states of topological materials such as 
topological superconductors and the superfluid $^3\mathrm{He}$-B phase\cite{Balian1963,Leggett1975,Schnyder2008}.
 In real TIs, the coupling between spin and orbital parts
has a more general form,
\begin{align}
H_0=a \alpha^{\mu}\Lambda_{\mu}^{\nu}p_{\nu}+M\beta, \label{H2}
\end{align}
where the configuration matrix $\Lambda_{\mu}^{\nu}$ defines the angle which locks spins and momenta,
and describes the rotation or the mirror operation in momentum space.
In such case, $\Lambda_{\mu}^{\nu}$ is the real symmetric matrix
satisfying $\Lambda^\nu_\mu\Lambda^\lambda_\mu=\delta^\nu_\lambda$.
The configuration of the spin-orbit interaction is determined by the lattice structure of the topological materials.
Experimentally, it is possible to confirm the structure of $\Lambda_{\mu}^{\nu}$ by the angle resolved photoemission spectroscopy in the presence of the Zeeman field. 
As shown in Appendix~\ref{AP1}, the shift of the Dirac point in the Zeeman field tells us details of $\Lambda_{\mu}^{\nu}$.

When we focus only on an isolated topological insulator, the physics of Eq.~(\ref{H2}) 
is the same as that of the simple Hamiltonian in Eq.~(\ref{Hamiltonian})
, because the Hamiltonian of Eq.~(\ref{H2}) can be connected with Eq.~(\ref{Hamiltonian}) by a unitary transformation~\cite{Liu2010}. 
However, when we consider a junction of two different topological insulators, physics happening near 
the junction interface depends on the choice of $\Lambda_{\mu}^{\nu}$ in the two topological insulators.
This is because there is no unitary transformation which 
transforms the two different $\Lambda_{\mu}^{\nu}$ into the Hamiltonian in Eq.~(\ref{Hamiltonian}) at the same time.
In the following, we study the properties of subgap states at the junction interface of two different topological insulators.
In the junction of the topological insulators, 
it is impossible to distinguish the two topological insulators in terms of $Z_2$ number
even if they have the different $\Lambda_{\mu}^{\nu}$.
This is because the $Z_2$ topological number is defined by all occupied states in the whole Brillouin zone of the TIs
and does not depend on the choice of the basis.
Therefore, it is necessary to define another topological number in the partial Brillouin zone to distinguish the two TIs.

\section{Topological numbers defined in the partial Brillouin zone}
To distinguish the two TIs, we need topological numbers defined in the partial Brillouin zone such as a two-dimensional plane at $p_x=0$ and a one-dimensional line at $p_x=p_y=0$.
The topological numbers defined in the partial Brillouin zone have a general property:
the summation of such a topological number over the whole Brillouin zone is zero.
To topologically distinguish the two TIs, therefore, we need to compare their topological numbers defined in the common partial Brillouin zone.
The topological numbers remain unchanged as far as perturbations 
do not mix the states in one partial Brillouin 
zone and those belonging to another partial Brillouin zone.
In the presence of potential disorder, strictly speaking, 
it would be difficult to define such independent partial Brillouin zones. 
In this paper, we consider the clean enough TIs to define the partial Brillouin zone.
The configuration of the spin-orbit interactions in the two TIs is represented by $\Lambda_{(i)}$ for $i=1,2$.
We fix $\Lambda_{(2)}$ at diag(1,1,1) without loss of generality 
because the relative configuration of momentum-spin locking is responsible for physics happening at the interface. 
In such a case, $\Lambda_{(1)}$ is identical to $\Lambda_{12}$. 

\subsection{Relative Chern number}\label{RC}
When $\Lambda_{(1)}$ is described as
\begin{align}
\Lambda_{(1)}=\begin{pmatrix}
\cos\phi&-\sin\phi&0\\
\sin\phi&\cos\phi&0\\
0&0&1
\end{pmatrix}\label{rotation},
\end{align}
we consider the Chern number in the two-dimensional Brillouin zone with $p_x=0$.
When the angle $\phi$ satisfies $\phi=\pi$, the Hamiltonian can be represented by
\begin{align}
H_{1(2)}=&\begin{pmatrix}
M\sigma^0&A'_{-(+)}\\
A'_{-(+)}&-M\sigma^0
\end{pmatrix},\label{HRC}\\
M=&m-b({p_y}^2+{p_z}^2),\\
A'_{\pm}=&a( p_z\sigma^z\pm p_y\sigma^y)
\end{align}
in either sides of the junction.
The discussion in the following can be applied to any two-dimensional partial Brillouin zones defined by $c_x p_x+c_y p_y=0$
 with real constants $c_\mu$.
Eq. (\ref{H2}) restricted by $c_x p_x+c_y p_y=0$ is transformed into Eq.~(\ref{HRC}) by changing the basis in spin space.
Especially for the angle $\phi=\pi$, the Hamiltonians of the two TIs can be unitary transformed into a block-diagonal form at the same time by use of $UW$ with
\begin{align}
U=\begin{pmatrix}
R^y(\pi/2)&0\\
0&R^y(\pi/2)
\end{pmatrix},\\
R^\mu(\theta)=\cos\frac{\theta}{2}\sigma^0-i\sin\frac{\theta}{2}\sigma^\mu,
\end{align} 
and
\begin{align}
W=\begin{pmatrix}
1&0&0&0\\
0&0&0&1\\
0&0&1&0\\
0&1&0&0
\end{pmatrix}.
\end{align}
The resultant Hamiltonians are 
\begin{align}
H_{1(2)}=&\begin{pmatrix}
M\sigma^z+A_{-(+)}&0\\
0&-M\sigma^z+A_{-(+)}
\end{pmatrix},\label{QHS}\\
A_{\pm}=&a( p_z\sigma^x\pm p_y\sigma^y).
\end{align}
Each blocked sector is equivalent to the quantum Hall Hamiltonian introduced by X.-L. Qi et al.\cite{Qi2006}.
The term $\pm M \sigma^z$ plays a role of 'fictitious magnetic field'.
The amplitude of the magnetic field is common in the two blocked sectors.
But the magnetic field in one sector points in the opposite direction to that in the other sector in Eq. (\ref{QHS}) as a 
consequence of the time-reversal symmetry of the original Hamiltonian in Eq. (\ref{H2}).
 When the $2\times 2$ Hamiltonian is given by
\begin{align}
h = s M\sigma^z+A_{\nu}
\end{align}
with $s=\pm 1$ and $\nu=\pm 1$, the Chern number is calculated to be $C= 1 \times \textrm{sgn}(s \nu)$.
When the Chern number of one blocked sector is $C=1$, that in the other is $C=-1$.
In the following, we represent a group of the quantum Hall Hamiltonian characterized by the first Chern number $C$ as $h(C)$.
Eq.~(\ref{QHS}) can be represented by the double quantum Hall systems without any interactions with each other as
\begin{align}
H_{1(2)}\in&\begin{pmatrix}
h\left(-(+)1\right)&0\\
0&h\left(+(-)1\right)
\end{pmatrix},\\
&M\sigma^z+A_+\in h(1),\label{Qimodel}
\end{align}
because unitary transformation of
\begin{align}
R^x(\pi)^\dagger(M\sigma^z+A_{-(+)})R^x(\pi)=-M\sigma^z+A_{+(-)}
\end{align}
does not change the Chern number $C$. 
Thus, the two blocked sectors have the opposite chiral edge modes to each other reflecting the opposite sign of the topological number.
In addition, when we focus one blocked sector, the Chern numbers in the two topological insulators have the opposite sign to each other.
Therefore we conclude that there are four chiral edge modes at the interface of the two TIs.

The argument to explain the appearance of the gapless states here is essentially the same as that in the previous work~\cite{Takahashi2011}.
Our explanation, however, does not need the presence of the mirror and the time-reversal symmetry. 
Thus the interface gapless state should be robust under the perturbation breaking the mirror symmetry.
For instance, the Zeeman field $H_\mu=B_\mu\sigma^\mu$ applied to the junction breaks the time-reversal symmetry and the mirror symmetry as shown in Appendix B.
We will numerically confirm the argument above in Sec.~IV.

Although they are fragile under deviating the rotation angle $\phi$ from $\pi$,
for $\phi\neq\pi$, it is impossible to transform the original Hamiltonian for the two TIs on both sides of junction 
into blocked Hamiltonian in Eq.~(\ref{QHS}) at the same time. 
The two TIs are not topologically distinct from each other.
As a result, the interface gapless states disappear for $\phi\neq\pi$.
We also numerically confirm this property in Sec. IV.

\subsection{Sato's winding number}\label{SWN}
Next, we consider Sato's winging number which is defined in the one-dimensional partial Brillouin zone with $p_x=p_y=0$,
when we choose $\Lambda_{(1)}$ as
\begin{align}
{\Lambda}_{(1)}=\begin{pmatrix}
c_{11} & c_{12} & 0\\
c_{21}& c_{22} & 0\\
0 & 0 & s
\end{pmatrix},\label{sato_matrix}
\end{align}
where the parameter $c_{ij}$ is chosen for $\Lambda_{(1)}$ to be the real symmetric matrix.
The Hamiltonian in Eq.~(\ref{H2}) is block diagonal in each spin space for $p_x=p_y=0$
irrespective of $c_{ij}$. Namely
\begin{align}
H_\uparrow=&
\begin{pmatrix} M_z & a s p_z \\
a s p_z & -M_z 
\end{pmatrix},\quad
H_\downarrow=
\begin{pmatrix}
M_z & -a s p_z \\
-a s p_z & -M_z 
\end{pmatrix},
\end{align}
with $M_z=m-bp_z^2$. 
For instance, eigen values of $H_\uparrow$ are $\pm \epsilon_{p_z}$ 
and eigen vectors are represented by
\begin{align}
 \begin{pmatrix}
 \cos(\theta_{p_z}/2) \\
 \sin(\theta_{p_z}/2)
 \end{pmatrix}\quad \textrm{and} \quad 
 \begin{pmatrix}-\sin(\theta_{p_z}/2) \\
 \cos(\theta_{p_z}/2)\end{pmatrix},
\end{align} for $\epsilon_{p_z}$ and  $-\epsilon_{p_z}$, respectively.
Here we define
\begin{align}
\epsilon_{p_z}=&\sqrt{M_z^2+ (ap_z)^2}, \\
\cos\theta_{p_z} =& \frac{M_z}{\epsilon_{p_z}}, \quad
\sin\theta_{p_z} = \frac{asp_z}{\epsilon_{p_z}}.
\end{align}
In the presence of the time-reversal symmetry, the eigen-vectors can be represented 
only by real quantities. 
In the presence of the band-inversion symmetry, Sato's winding number 
can be defined at $p_x=p_y=0$ as
\begin{align}
\mathcal{W}(\uparrow,s)=\frac{1}{2\pi} \int_{-\pi}^\pi dp_z \partial_{p_z} \theta_{p_z}.
\end{align}
To estimate the topological number, we use tight-binding representation 
of the Hamiltonian,
\begin{align}
M_z=&m-bp_z^2 \to m - 2t( 1- \cos p_z), \\
ap_z & \to a_0 t \sin p_z,
\end{align}
with $t>0$, $m-4t<0$, and $|a_0| \ll 1$ is a dimensionless constant.
We find that 
\begin{align}
\mathcal{W}(\uparrow,s)=\textrm{sgn}(s), \quad \mathcal{W}(\downarrow,s)=-\textrm{sgn}(s).
\end{align}
Therefore, when the two TIs in the junction have opposite sign of $s$, 
the topological gapless interface states are guaranteed by the difference of Sato's winding number in two spin space.

In the absence of the time-reversal symmetry under the Zeeman field, it is also possible to 
discuss the robustness of the gapless states by using another topological number~\cite{Mizushima2012}.
To apply their argument, we first transform Eq.~(\ref{H2}) under the Zeeman field as
\begin{align}
&U_{\textrm{TH}}
\begin{pmatrix}
 M \sigma^0 + \boldsymbol{h} \cdot \boldsymbol{\sigma} & a {\sigma}^\mu\Lambda_\mu^\nu p_\nu \\
a {\sigma}^\mu\Lambda_\mu^\nu p_\nu & - M \sigma^0 + \boldsymbol{h} \cdot \boldsymbol{\sigma}
\end{pmatrix}
U^\dagger_{\textrm{TH}} \nonumber\\ 
&= 
\begin{pmatrix}
 -M \sigma^0 + \boldsymbol{h} \cdot \boldsymbol{\sigma} & a {\sigma}^\mu\Lambda_\mu^\nu p_\nu i\sigma^y\\
-i\sigma^y a {\sigma}^\mu\Lambda_\mu^\nu p_\nu &  M \sigma^0 - \boldsymbol{h} \cdot \boldsymbol{\sigma}^\ast
\end{pmatrix}=H_{He},\label{he3-b}\\
&U_{\textrm{TH}}=\begin{pmatrix}
 0 & 1 \\ -i\sigma^y & 0 
\end{pmatrix}.
\end{align}
The right-hand side of Eq.~(\ref{he3-b}) is the Hamiltonian of the $^3$He-B phase under the Zeeman field $\boldsymbol{h}$.
Under the condition in Eq.~(\ref{sato_matrix}) with $p_x=p_y=0$ and $\boldsymbol{h}=(h_x,h_y,0)$, the Hamiltonian satisfies the relation as
\begin{align}
&\{ \Gamma,  H_{He}\}_+ =0, \quad \Gamma = \mathcal{C} \mathcal{T}{\Pi},\\
&\mathcal{C}=\begin{pmatrix} 0 & \mathcal{K} \\ \mathcal{K} & 0 \end{pmatrix}, \quad
\mathcal{T}=\begin{pmatrix} i\sigma^y \mathcal{K}&0 \\  0 & i\sigma^y \mathcal{K} \end{pmatrix},\\
&{\Pi}=\begin{pmatrix} i\sigma^z  & 0 \\  0 & -i\sigma^z \end{pmatrix},
\end{align}
where $\mathcal{K}$ means the complex conjugation, $\Pi$ represents the $\pi$ rotation in the $xy$ plane in spin space, 
$\mathcal{T}$ means the time-reversal operation, and $\mathcal{C}$ represents the band-inversion plus the complex conjugation.
The Hamiltonian $H_{He}$ is transformed as
\begin{align}
U_{\Gamma} H_{He} U_{\Gamma}^\dagger = 
\begin{pmatrix}0 & q \\ q^\dagger& 0 \end{pmatrix},\; 
U_\Gamma=\frac{q}{\sqrt{2}}\begin{pmatrix} \sigma^0 & -i \sigma^x \\ -i \sigma^x & \sigma^0 \end{pmatrix}.
\end{align}
A topological number is defined by
\begin{align}
W(s) =& - \frac{1}{4\pi i} \int dk_z \textrm{Tr}[ \Gamma H_{He}^{-1} \partial_{k_z} H_{He} ],\\
 =&
\frac{1}{2\pi} \textrm{Im} \int dk_z \partial_{k_z} \log( \textrm{det} q).
\end{align}
We find $W(s)= 2s$ for $|\boldsymbol{h}|< m$. When $\Lambda_{(1)}$ in Eq.~(\ref{sato_matrix}) represents 
the inversion of $p_z$ axis (i.e., $s=-1$), two topological insulators are topologically 
distinct from each other even in the presence of the Zeeman field parallel to the plane.
Four subgap states appear at zero energy and at $p_x=p_y=0$ because of $|W(1)-W(-1)|=4$.
It is still unclear the effects of the Zeeman field perpendicular to the junction plane $h_z$ and those of the 
perturbation breaking the band-inversion symmetry on the interface gapless states. We will numerically 
check these issues 
in Sec.~IV.

Finally, we discuss the stability of the gapless states characterized by Sato's number.
The gapless interface states are insensitive to choice of $c_{ij}$ in Eq.~(\ref{sato_matrix}) because 
 Sato's number is defined in one-dimensional Brillouin zone at $p_x=p_y=0$.
Therefore, the experimental realization of the gapless states characterized by Sato's number is 
much easier than that of of the gapless states characterized by the Chern number in Sec.~IIIA.

\section{Numerical simulation on tight-binding model}
To confirm the conclusions in the topological analysis and 
to check the robustness of the interface gapless states under perturbations, we also perform the numerical 
calculation on the tight-binding model.
We describe the topological insulator by using the two-band model as
\begin{align}
&H_{0}(\boldsymbol{k};j,j') = 
%\sum_{j,j'}\sum_{\boldsymbol{k}}
%\left[\begin{array}{c}
%\tilde{c}_{\boldsymbol{k},j',1}\\
%\tilde{c}_{\boldsymbol{k},j',2}\end{array}
%\right]^\dagger
\begin{pmatrix}
M{\sigma}^0 & h_{\textrm{so}}\\
%\boldsymbol{A}\cdot {\boldsymbol{\sigma}} \\
%\boldsymbol{A}\cdot {\boldsymbol{\sigma}} 
h_{\textrm{so}} & -M{\sigma}^0
\end{pmatrix},\label{tbh0}\\
%\nonumber\\
%&\times
%\begin{pma
%\tilde{c}_{\boldsymbol{k},j,1}\\ \tilde{c}_{\boldsymbol{k},j,2}\end{array}
%\right],\\
%\tilde{c}_{\boldsymbol{k},j,\nu}=& \left[\begin{array}{c}
%c_{\boldsymbol{k},j,\nu,\uparrow}\\
%c_{\boldsymbol{k},j,\nu,\downarrow}\end{array}\right],\\
&M = (m +2b_2 (\cos(k_xc_0) + \cos(k_yc_0)-2)))\delta_{j,j'}\nonumber\\
&\qquad- 2b_1\delta_{j,j'}+ b_1 (\delta_{j,j'+1} + \delta_{j,j'-1})
, \\
&h_{\textrm{so}}=
a_2 \sigma^\nu\left[ \Lambda_\nu^x\sin(k_xc_0) +\Lambda_\nu^y\sin(k_yc_0)\right]\delta_{j,j'} \nonumber\\
&-ia_1 \sigma^\nu \Lambda_\nu^z(\delta_{j,j'+1}-\delta_{j,j'-1}),
\end{align}
where $\boldsymbol{k}=(k_x,k_y)$ is the two-dimensional momentum parallel to the surface and $j$ represents the position in the $z$-direction.
We apply the periodic boundary condition in the $xy$-plane and 
the hard wall boundary condition in the $z$-axis.
We choose $a_1=7.86m/c_0$, $a_2=14.6m/c_0$,  $b_1=3.57\times10m/c_0^2$, $b_2=2.02\times10^2m/c_0^2$ 
with the lattice constant $c_0$ being $5\; [\mathrm{\AA}]$ in both sides 
taking into account the band structures of $\mathrm{Bi_2Se_3}$~\cite{H.Zhang2009}.   
In the simulation, the length of the each topological insulator in the $z$-direction is 400 $c_0$.

We also study the spin polarization of the interface gapless states 
and calculate the spin expectation value by use of the two spin operators as,
\begin{align}
S^\mu=&\begin{pmatrix}
\sigma^\mu&0\\
0&\sigma^\mu
\end{pmatrix},\label{S1}\\
\tilde{S}^\mu=&\begin{pmatrix}
0&\sigma^\mu\\
\sigma^\mu&0
\end{pmatrix},\label{S2}
\end{align}
where $S^\mu$ is the ordinary spin operator and $\tilde{S}^\mu$ is derived from the basis of the wave function in the topological insulator\cite{Liu2010}.
We utilize the two operators depending on the basis of the interface states.
In the following, we calculate $S^\mu$ according to the spin of the ordinary surface state on TIs in Eq.~(\ref{Hamiltonian})
when the energy dispersion is shown along $k_\mu$. 

To check the robustness of the interface zero-energy states, we also consider 
two types of perturbations breaking the band-inversion and time-reversal symmetries in addition to Eq.~(\ref{tbh0})
\begin{align}
&H_{\textrm{BI}}(j,j') =  1_{4 \times 4}\nonumber\\
&\times\left\{\begin{array}{cc} \mu_{(1)}\delta_{j,j'}
+b_{(1)}(\delta_{j,j'+1}+\delta_{j,j'-1}) & \textrm{for}\; j<0 \\
\mu_{(2)}\delta_{j,j'}
+b_{(2)}(\delta_{j,j'+1}+\delta_{j,j'-1}) & \textrm{for}\; j>0\end{array} \right. \label{hbi}\\
&H_{\textrm{TR}}=\begin{pmatrix}
\boldsymbol{h} \cdot \boldsymbol{\sigma} & 0\\
0 & \boldsymbol{h} \cdot \boldsymbol{\sigma} 
\end{pmatrix}.\label{htr}
\end{align}
 Eq.~(\ref{hbi}) represents the shift of the chemical potential and 
the modification of the band width introduced differently 
in the two TIs. Such a perturbation breaks the band-inversion symmetry.
The choice of $\mu_{(1)}=\mu_{(2)}$ and $b_{(1)}=b_{(2)}$ also breaks the band-inversion symmetry. 
Such a perturbation, however, does not affect the difference of the topological numbers in the two TIs because
they are represented as the global energy shift proportional to $1_{4 \times 4}$.  
Therefore, such a perturbation does not affect the difference of topological numbers in the two topological insulators.
Thus we choose $\mu_{(1)}\neq\mu_{(2)}$ and $b_{(1)}\neq b_{(2)}$ in the numerical simulation.
It is impossible to describe such perturbation as a common energy shift in the two TIs. 
Eq.~(\ref{htr}) represents the uniform Zeeman potential which breaks the time-reversal 
symmetry. 
Although the Zeeman potential is band-inversion asymmetric, we can extract the effect of 
the time-reversal breaking perturbation.
In the simulation, we introduce Eq.~(\ref{htr}) commonly in the two TIs.
% by use of the $H_{\mathrm{TR}}$ commonly in the two TIs. 
In such a case, effects of band-inversion asymmetry on the difference of the topological numbers in the two TIs 
can be eliminated.
% in the difference of the topological numbers in the two TIs 
%when Eq.~(\ref{htr}) is introduced commonly in the two TIs. 
Therefore, it is possible to study the effects of the breakdown of the band-inversion symmetry and 
those of the time-reversal symmetry 
independently by introducing Eqs. (\ref{hbi}) and (\ref{htr}), respectively.

In what follows, we show the numerical results of the energy dispersion and spin polarization in the junction of two TIs
with two different $\Lambda$.
Some of the spectra of the interface states are also analytically calculated within the quasiclassical approximation as shown in Appendix \ref{quasi}.
We also discuss effects of disorder on the interface states. In the presence of disorder, 
the partial Brillouin zone specified by a wavenumber is not well defined. 
We will show that the interface gapless states are fragile under the disordered potential in Appendix \ref{AP4}.

\subsection{Rotation within the $xy$-plane}
When we choose 
\begin{align}
{\Lambda}_{(1)}=\begin{pmatrix}
\cos\phi & -\sin\phi & 0 \\
\sin\phi& \cos\phi & 0\\
0 & 0 & 1
\end{pmatrix}, \label{mirror}
\end{align}
the spin-orbit interaction in Eq.~(\ref{mirror}) describes only the continuous rotation 
of spin-momenta locking angle $\phi$ within the $xy$ plane. 
The spin-momenta locking in the $z$ axis is common in the two topological insulators.
In this case, the interface gapless state is predicted only at $\phi=\pi$ from the analysis of the relative Chern number in Sec.~\ref{RC}.
\begin{figure}[htbp]
  \subfigure{\includegraphics*[height=42mm]{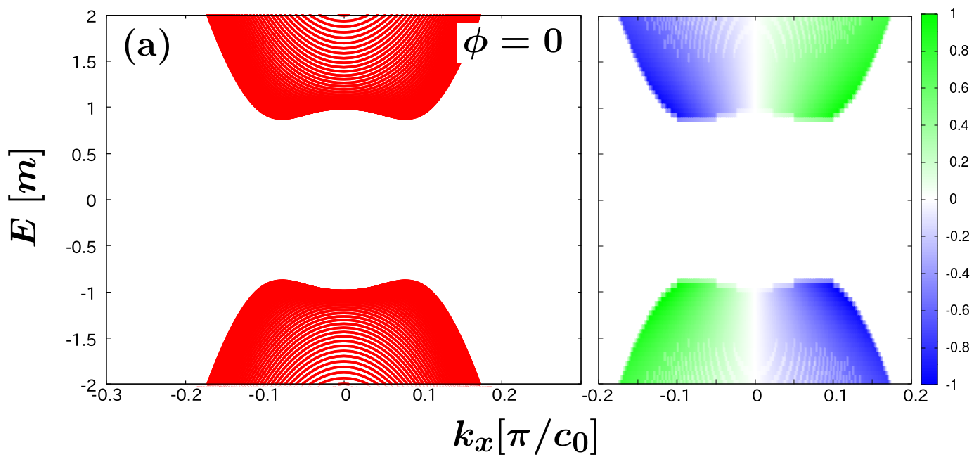}\label{fig:0}}
  \subfigure{\includegraphics*[height=42mm]{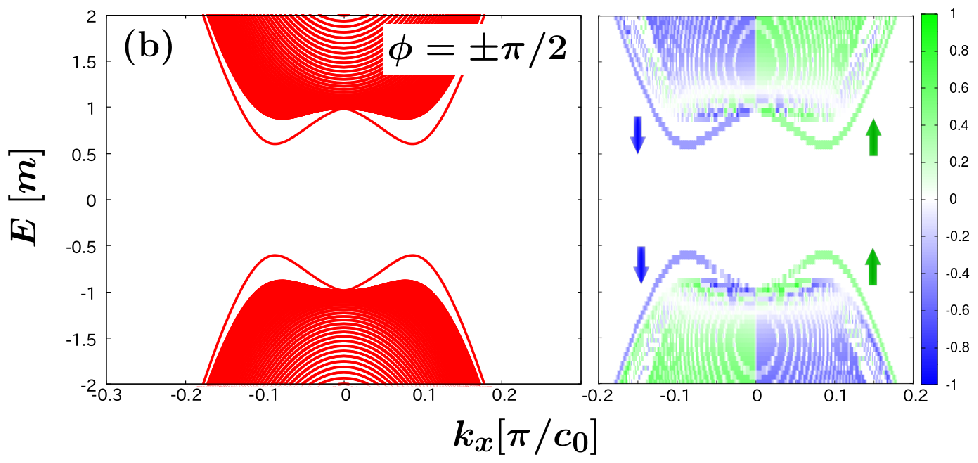}\label{fig:pi2}}
   \subfigure{\includegraphics*[height=42mm]{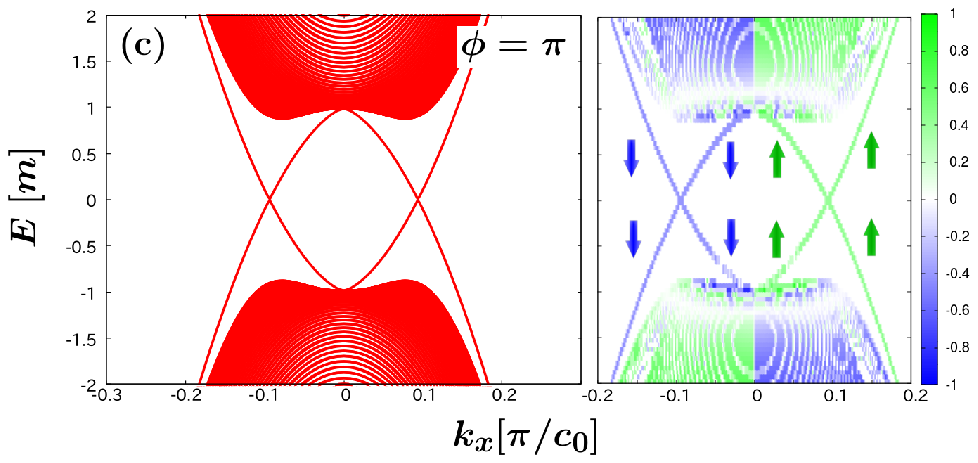}\label{fig:pi}}
\caption{The energy dispersions(left figures) and the spin configurations(right figures) of the interface states along $k_x$ are shown for Eq.~(\ref{mirror}).
The dispersions are plotted on the $k_x$ only because the dispersions have the rotational symmetry in two-dimensional momentum $(k_x,k_y)$.
We choose $\phi=0,\;\pi/2$, and $\pi$ in (a), (b), and (c), respectively in Eq.~(\ref{mirror}).
Only for $\phi=\pi$, the gapless interface state appears in the gap of the bulk state.
 }\label{fig1}
\end{figure}
In Fig.~\ref{fig1}, we show numerical results of the dispersion relations for
several choices of $\phi$.
The junction under consideration has the two surfaces of topological insulators which host 
the gapless surface states. We delete the contribution from such surface states.
In the numerical results, we plot the energy eigenvalues of the states in bulk and those localized at the junction 
interface.
The dispersions are plotted as a function of $k_x$ only
because the numerical results are isotropic in momentum space. 
It is evident that the interface subgap state is absent at $\phi=0$ 
because the resulting relation ${\Lambda}_{(2)}={\Lambda}_{(1)}$ means 
the junction of two identical topological insulators.
The gap of the interface state decreases with increasing $\phi$ to $\pi$ 
as shown in Figs.~\ref{fig:0} and \ref{fig:pi2}.
The momenta at the minima in the upper band and the maxima in the lower band have the 
ring-shaped in two dimensional Brillouin zone.
At $\phi=\pi$, such minima and maxima touch each other as shown in Fig.\ref{fig:pi}.
The interface states become gapless and form a ring-shaped Fermi surface.
The subgap spectra obtained within the quasiclassical
approximation $a^2\ll mb$ in Appendix C show the property consistent with the numerical results, 
\begin{align}
E=a\sqrt{{p_F}^2-|\boldsymbol{k}|^2\sin^2(\phi/2)},
\end{align}
where $p_F=\sqrt{m/b}$ is a real value satisfying $p_F \geq |\boldsymbol{k}|$ with $\boldsymbol{k}=(k_x, k_y)$.
The zero-energy state is possible only at $|\boldsymbol{k}|=p_F$ and $\phi=\pi$.
The spin polarization of the interface state is calculated by use of $S^x$ in Eq.~(\ref{S1}) and 
shown with the dispersion. The results suggest the Kramers degeneracy in the subgap states 
in the presence of the time-reversal symmetry.
The dispersion along radial momentum represents a linear dispersion from the ring-shaped zeros.

The topological argument in Sec.~IIIA, the numerical results, and the analytical 
expression within the quasiclassical approximation suggest that the gapless states appear only at $\phi=\pi$.
Therefore the delicate material tuning is necessary for having the
ring-shaped zero-energy state in experiment. 
A small perturbation modifying the locking angle $\phi$ open the gap.

Next we check the robustness of the gapless interface state at $\phi=\pi$ by introducing the 
Zeeman field in Eq.~(\ref{htr}) which breaks time-reversal symmetry.
\begin{figure}[htbp]
    \subfigure{\includegraphics*[height=45mm]{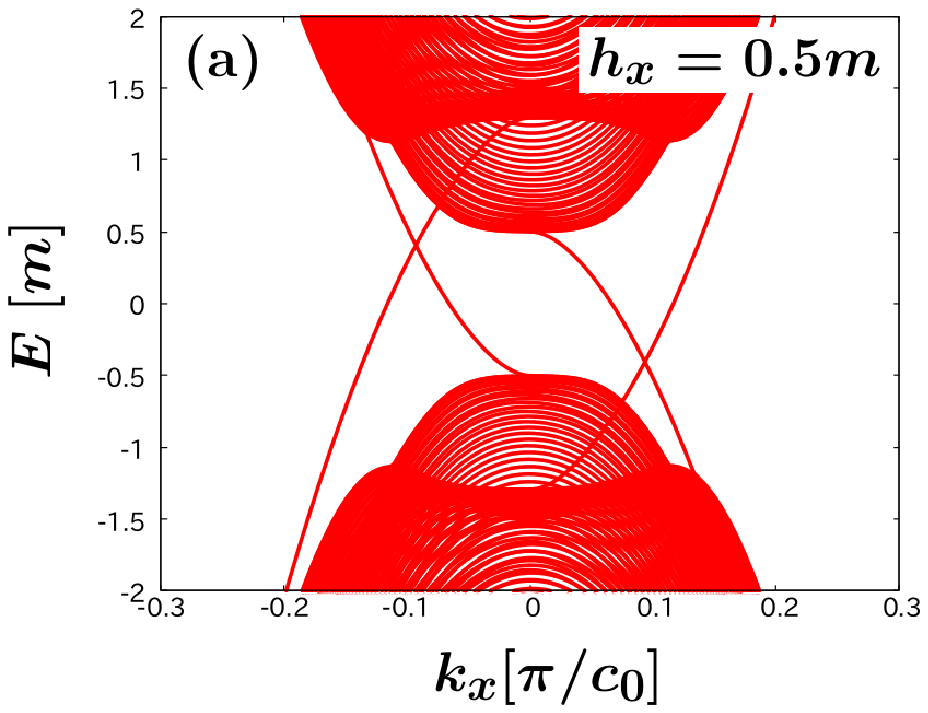}\label{fig:inp}}
    \subfigure{\includegraphics*[height=45mm]{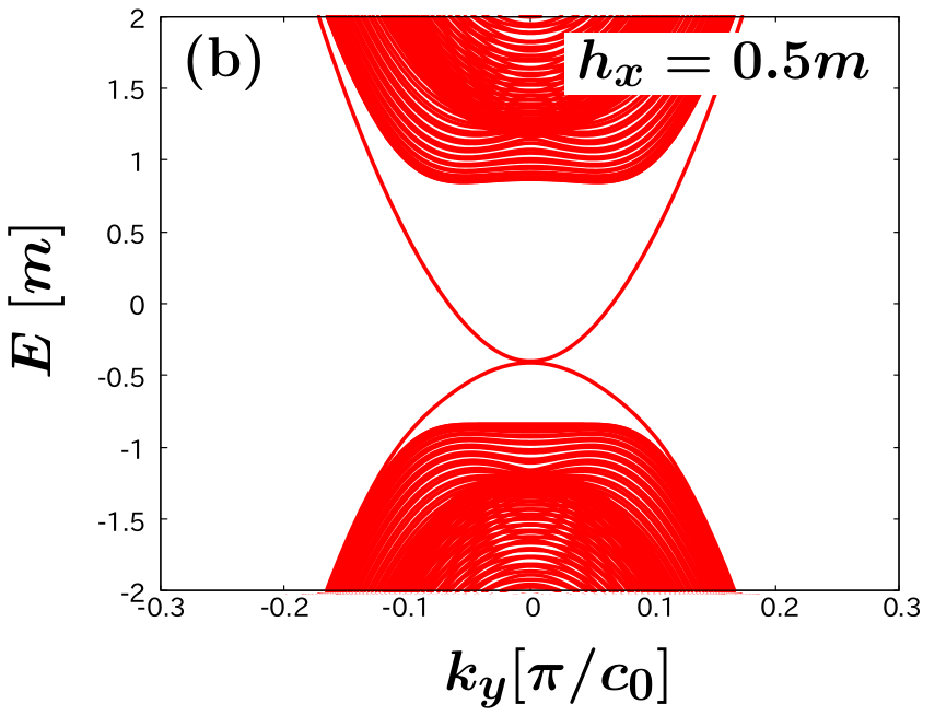}\label{fig:inp2}}
   \subfigure{\includegraphics*[height=45mm]{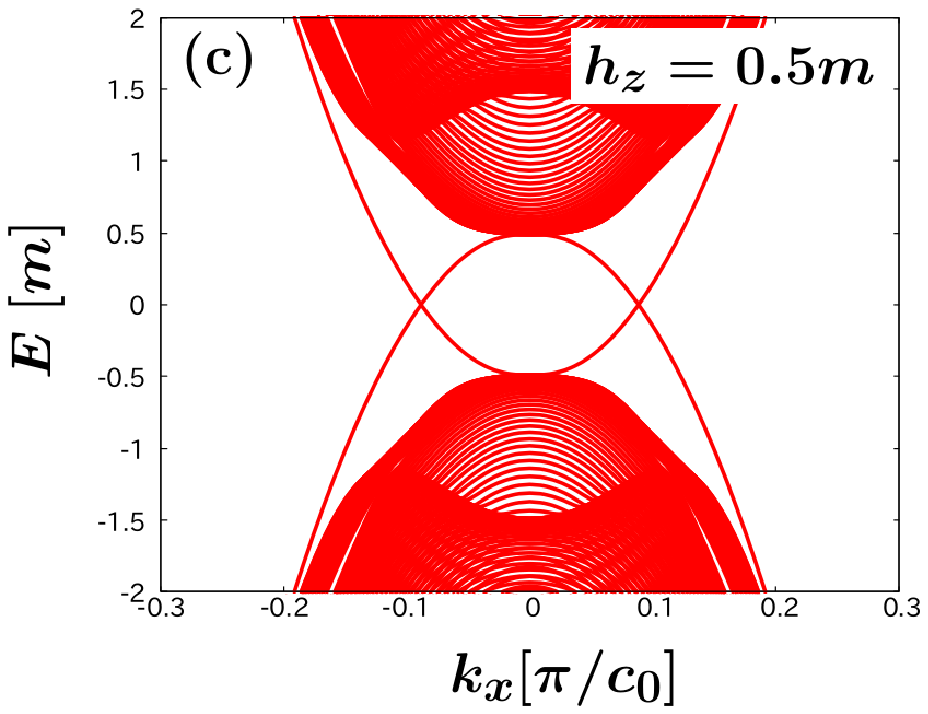}\label{fig:per}}
\caption{The energy dispersions of interface states along $k_x$ in the presence of the Zeeman field are shown for 
Eq.~(\ref{mirror}) with $\phi=\pi$.
The figure (a) and (b) are calculated along $k_x$ and $k_y$ axes respectively in the presence of the Zeeman field of $h_{x}=0.5m$ which is parallel to the interface.
The figure (a) is shown with $k_y=0$ along $k_x$.
In the fig.~(b), the dispersion is drawn along $k_y$ with $k_x$ on the point at which the upper and lower surface band are touching
in the positive $k_x$.
The figure (c) is calculated in the presence of the Zeeman field of $h_{z} =0.5m$ which is perpendicular to the interface.
 }\label{fig2}
\end{figure}
In Fig.~\ref{fig2}, we consider the Zeeman field parallel to the interface plane with $h_x=0.5m$ and $h_y=h_z=0$ in (a) and (b), 
and that perpendicular to the interface plane with $h_z=0.5m$ and $h_x=h_y=0$ in (c).
The interface gapless state shows the robustness in the presence of the Zeeman field 
reflecting the absence of mirror and time-reversal symmetries in the definition of the relative Chern number
in Sec. IIIA.
The robustness is attributed to the fact that the relative Chern number is insensitive to the Zeeman field.
The previous paper\cite{Takahashi2011} has concluded that the mirror symmetry protects the gapless states.
However, our numerical calculation shows that the gapless states are robust to the Zeeman field breaking mirror symmetry as discussed in Appendix \ref{APmirror}. 
\begin{figure}[htbp]
    \subfigure{\includegraphics*[height=45mm]{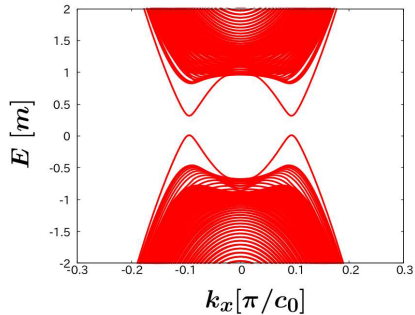}\label{fig:inpbi}}
\caption{The energy dispersion of interface states along $k_x$ is shown for 
Eq.~(\ref{mirror}) with $\phi=\pi$.
The dispersion is calculated in the presence of the band asymmetry as 
$\mu_{(1)}=-\mu_{(2)}=0.1m$ and $b_{(1)}=0.1b_1$.
 }
\end{figure}

In Fig.~\ref{fig:inpbi}, we calculate the energy dispersion at $\phi=\pi$ in the presence of 
the band asymmetrical perturbation in Eq.~(\ref{hbi}), where we 
introduce the difference of the band asymmetric term $\mu_{(1)}=-\mu_{(2)}=0.1m$, $b_{(1)}=0.1b_1$ and $b_{(2)}=0$ 
in Eq.~(\ref{hbi}).
It is clear from Fig. \ref{fig:inpbi} that the band asymmetry removes the gapless interface state.

\subsection{Inversion of the $z$-axis}\label{NRI}
Next we choose 
\begin{align}
{\Lambda}_{(1)}=\begin{pmatrix}
\cos\phi & -\sin\phi & 0\\
\sin\phi& \cos\phi & 0\\
0 & 0 & -1
\end{pmatrix} .\label{inversion}
\end{align}
The spin-orbit interaction in Eq.~(\ref{inversion}) describes the inversion in the $z$-axis 
plus the continuous rotation of spin-momenta locking angle $\phi$ within the $xy$ plane.
The inversion in the $z$-axis leads to the difference of Sato's winding number
in each spin sector~\cite{SatoM2010,Yada2011,SatoM2011} as discussed in Sec.~\ref{SWN}.
Because Sato's winding number can be defined in the one-dimensional Brillouin zone, 
it is independent of terms vanishing in the limit of $\boldsymbol{k}=(k_x,k_y)=0$.

\begin{figure}[htbp]
   \subfigure{\includegraphics*[height=42mm]{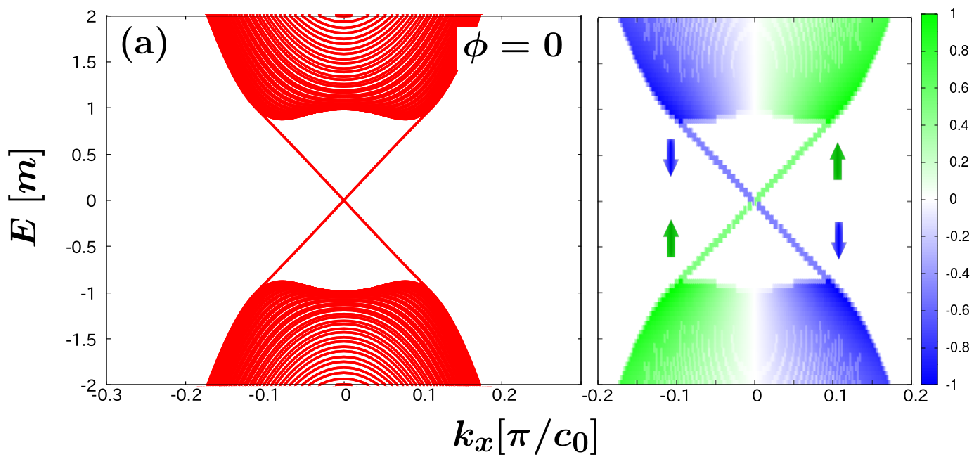}\label{fig:az}}
     \subfigure{\includegraphics*[height=42mm]{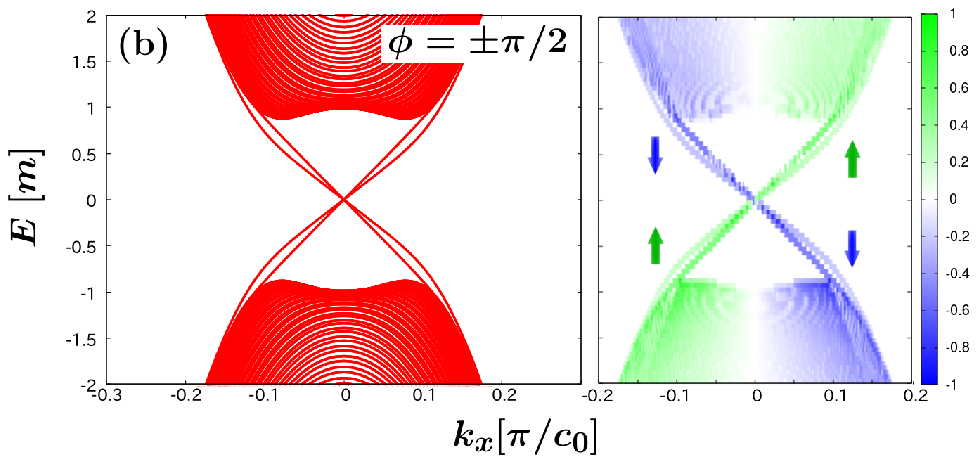}\label{fig:az+hpi}}
   \subfigure{\includegraphics*[height=42mm]{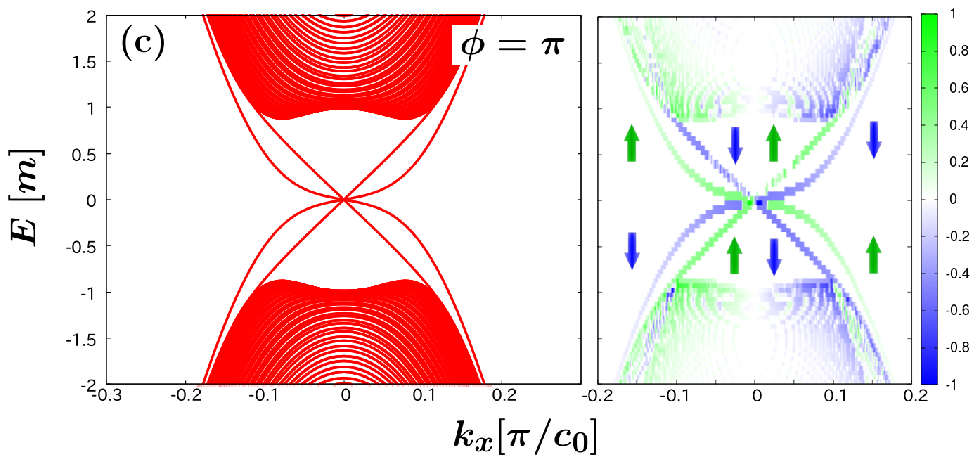}\label{fig:az+pi}}
\caption{The energy dispersions and the spin configurations of interface states along $k_x$ are shown for 
Eq.~(\ref{inversion}).
We choose $\phi=0,\;\pi/2$,and $\pi$ in (a), (b), and (c), respectively.
The gapless interface state appears for all $\phi$, which shows the robustness of the gapless state
under changing the rotation angle $\phi$.
 }\label{fig4}
\end{figure}

The topological numbers are calculated as 
\begin{align}
\mathcal{W}_{(1),\uparrow}=-1, \quad \mathcal{W}_{(1),\downarrow}=1
\end{align}
for $z<0$ and
\begin{align}
\mathcal{W}_{(2),\uparrow}=1, \quad \mathcal{W}_{(2),\downarrow}=-1
\end{align}
for $z>0$.
The bulk-boundary correspondence suggests that the number of 
zero-energy states for spin-up space is equal to 
$|\mathcal{W}_{(1),\uparrow}-\mathcal{W}_{(2),\uparrow}|=2$ 
 and that for the spin-down one is 
$|\mathcal{W}_{(1),\downarrow}-\mathcal{W}_{(2),\downarrow}|=2$.
As a consequence, four states are degenerate at zero energy and 
at $\boldsymbol{k}=0$. 

The numerical results in Fig.~\ref{fig4}(a) show the doubly degenerate Dirac cones at the interface according to the topological analysis.
The degeneracy at $\boldsymbol{k}=0$ is stable under the rotation of $\phi$ as shown in Figs.~\ref{fig4}(b) and (c).
The result of the quasiclassical calculation for Eq. (\ref{inversion}) in Appendix C shows similar spectra.
For $\phi=\pi$, a pair of dispersion branches show the linear dispersion, whereas the other pair have the cubic dispersion.
The spin polarization on the right panel of Fig. \ref{fig4}(c) suggests the cubic dispersion, 
where the Kramers pairs are always degenerate in the presence of the time-reversal symmetry. 
The spin polarization calculated by use of $\tilde{S}^x$ in Eq.~(\ref{S2}) shows the helical spin configuration.
We note that the choice of $\phi=\pi$ corresponds to the inversion for the three axes.

\begin{figure}[htbp]
    \subfigure{\includegraphics*[height=45mm]{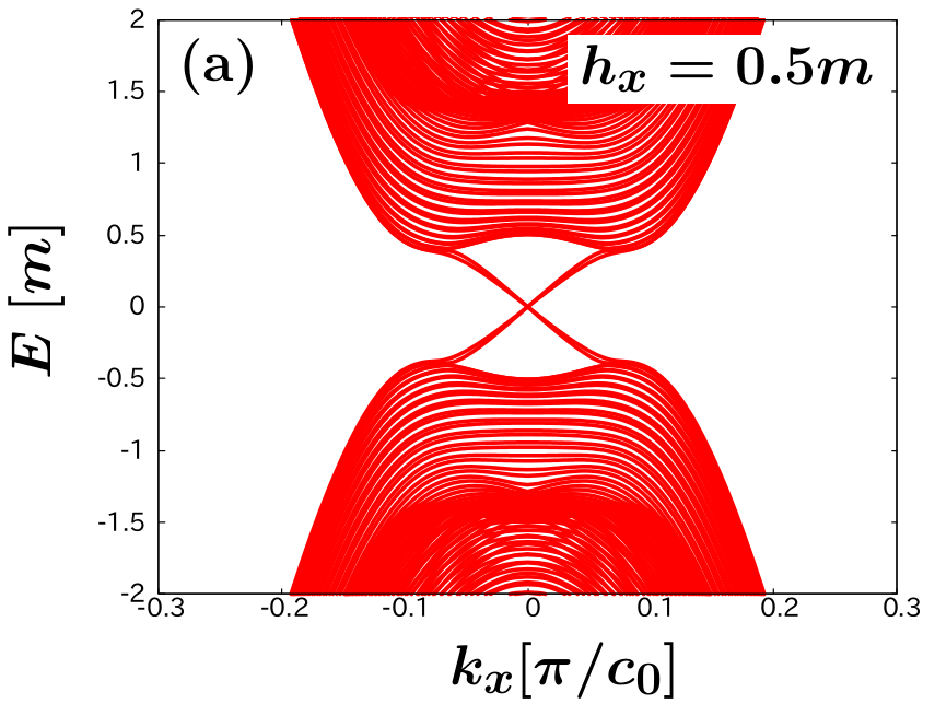}\label{fig:invinp}}
     \subfigure{\includegraphics*[height=45mm]{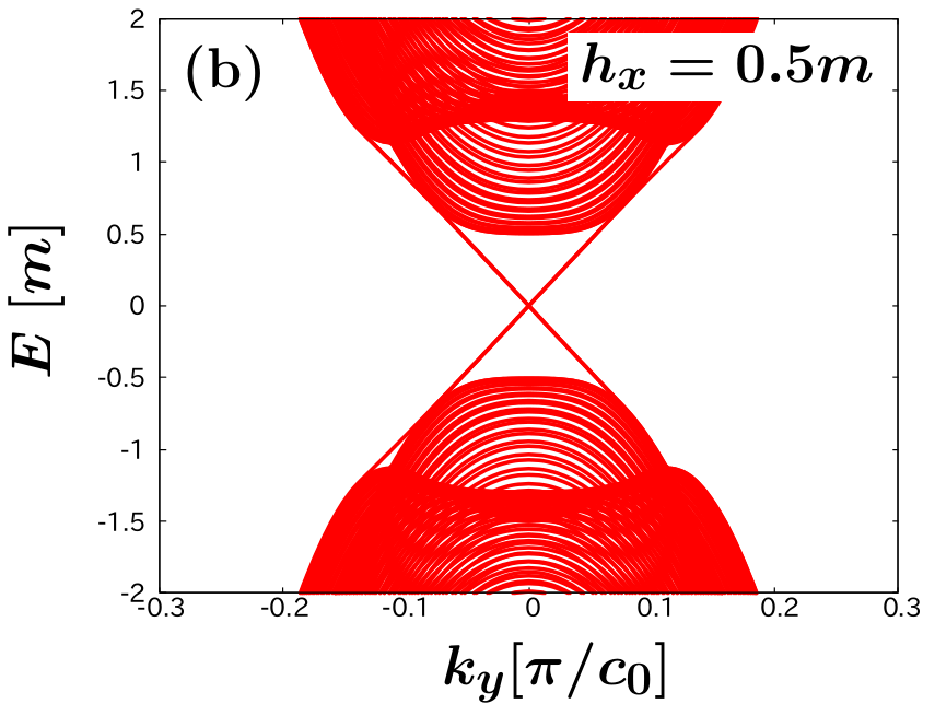}\label{fig:invinp2}}
     \subfigure{\includegraphics*[height=45mm]{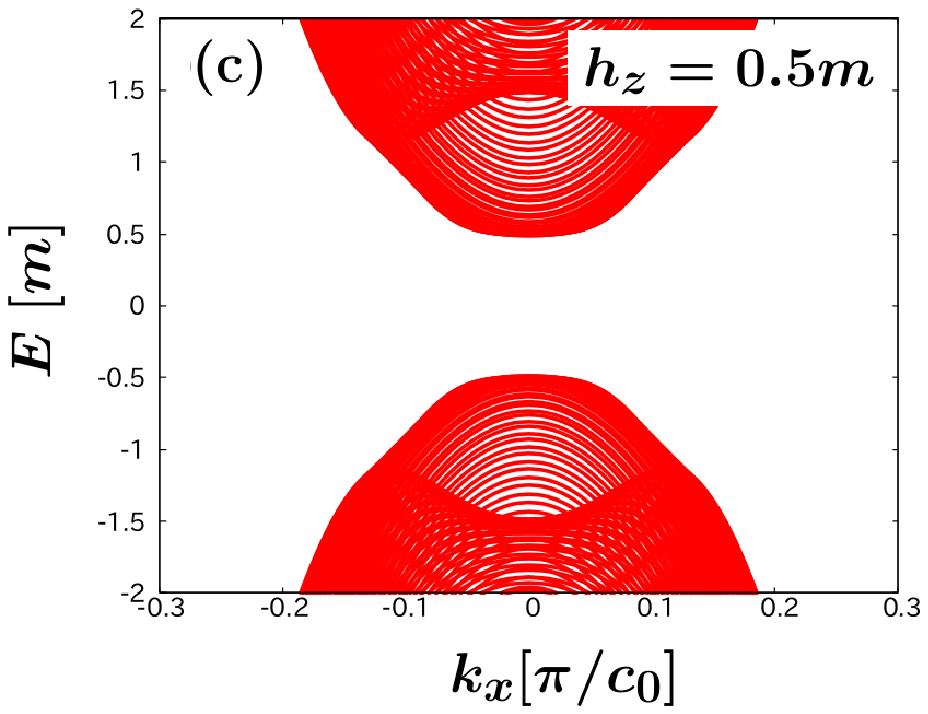}\label{fig:invper}}
\caption{The energy dispersions of interface states along $k_x$ in the Zeeman field are shown for 
Eq.~(\ref{inversion}) at $\phi=0$.
The Zeeman field is parallel to the $xy$ plane $h_{x}=0.5m$ in (a) and (b), and is perpendicular to the $xy$ plane 
$h_{z} =0.5m$ in (c).
 }
\end{figure}

We check the robustness of the zero-energy states under the perturbations breaking the 
time-reversal symmetry in Eq.~(\ref{htr}). 
In Figs.~\ref{fig:invinp} and (b), we show the dispersion of the interface states in the presence of the 
Zeeman fields parallel to the interface
 with $h_x=0.5m$, $h_y=0$ and $h_z=0$ in (a) and 
(b), and the Zeeman field perpendicular to the interface 
 with $h_x=h_y=0$ and $h_z=0.5m$ in (c). The dispersion is plotted as a function of $k_x$ in (a) and $k_y$ in (b). 
The results in Fig.~\ref{fig:invinp} show that the two Dirac nodes 
stay at the zero-energy and at $\boldsymbol{k}=0$ even in the in-plane Zeeman field.
When we apply the Zeeman field along the $z$-axis, on the other hand, the interface zero-energy states 
vanish as shown in Fig.~\ref{fig:invper}. 
Therefore the robustness of zero energy states 
depends on the direction of the Zeeman field. 
The zero-energy states are fragile(robust) against 
the Zeeman field perpendicular (parallel) to the interface. 
The magnetic anisotoropy is consistent with that found in the topological gapless states at the surface of $^3$He-B 
phase~\cite{Nagato2009,Chung2009,Nagato2011} as discussed in Sec.~IIIB. 

\begin{figure}[htbp]
     \subfigure{\includegraphics*[height=45mm]{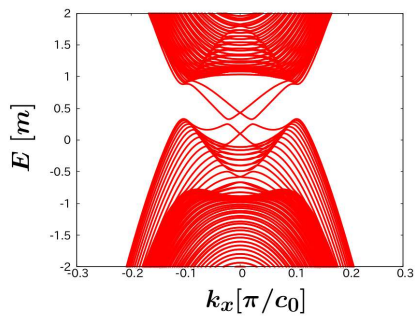}}
\caption{The energy dispersion of interface states along $k_x$ is shown for 
Eq.~(\ref{inversion}) at $\phi=0$.
The dispersion is calculated in the presence of chemical potentials of $\mu_{(1)}= -\mu_{(2)} =0.2m$ and 
the band asymmetry $b_{(1)}=0.25b_1$ and $b_{(2)}=0$.
 }\label{fig:dmu}
\end{figure}
In Fig.~\ref{fig:dmu}, we discuss effects of the band asymmetric perturbation
in Eq.~(\ref{hbi}) with $\mu_{(1)}=-\mu_{(2)}=-0.2m$, $b_{(1)}=0.25b_1$ and $b_{(2)}=0$. 
The two Dirac cones are lifted off from each other and the gap opens in the presence of the band asymmetric perturbation.

\subsection{Remaining configurations}
\begin{figure}[htbp]
    \subfigure{\includegraphics*[height=45mm]{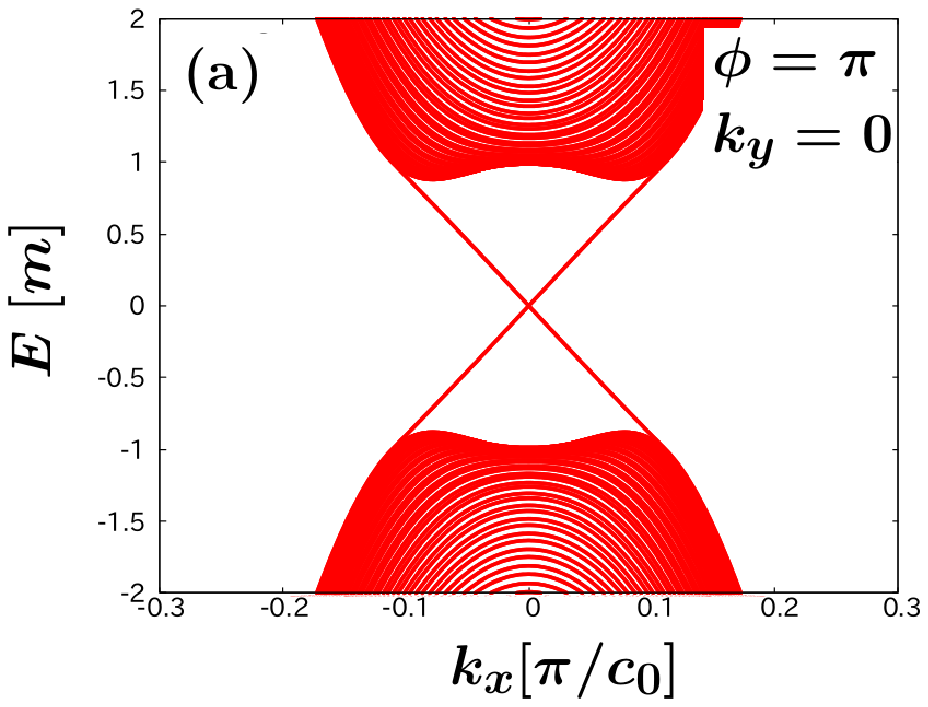}\label{fig:invxz}}
     \subfigure{\includegraphics*[height=45mm]{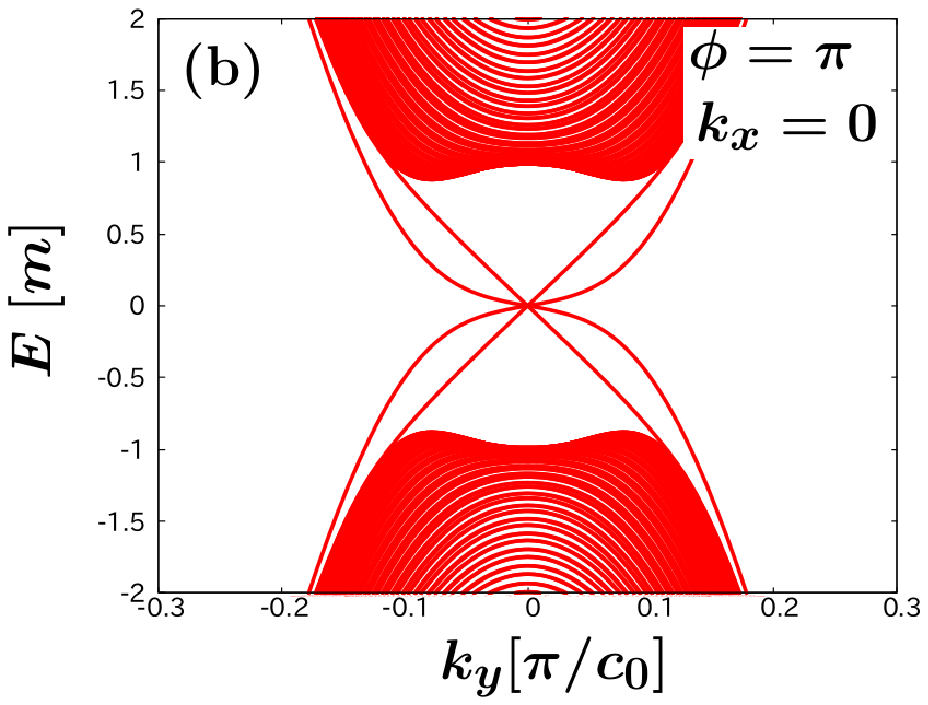}\label{fig:invyz}}
\caption{The energy dispersions of interface states in Eq.~(\ref{invx}) with $s=1$ and $\phi=\pi$
are shown along $k_x$ and $k_y$ in (a) and (b), respectively.
 }\label{inv}
\end{figure}

Finally we choose 
\begin{align}
{\Lambda}_{(1)}=\begin{pmatrix}
 s & 0 & 0\\
0& \cos\phi & -\sin\phi \\
0& \sin\phi& \cos\phi 
\end{pmatrix}, \label{invx}
\end{align}
with $s=\pm1$. 
For $s=1$, the transformation represents only the continuous rotation within the $yz$-plane.
The gapless states appear only at $\phi=\pi$. 
The dispersion becomes anisotropic in the two-dimensional Brillouin zone as shown in 
Figs.~\ref{fig:invxz} and \ref{fig:invyz}
because $x$ and $y$ axes are 
no longer equivalent to each other. 
The energy dispersions along $k_x$- and $k_y$-axes are the same as 
the dispersion with $\phi=0$ and that with $\phi=\pi$ in Eq.~(\ref{inversion}), respectively. 
Sato's winding number well characterizes the interface gapless states in Fig. \ref{inv} as well as 
those in Fig. \ref{fig4}. 
The gapless states in Fig \ref{inv} guaranteed by Sato's winding number
also have the robustness of the Zeeman field parallel to the interface 
and are fragile against that perpendicular to the interface similar to Fig. 5.
On the other hand, the relative Chern numbers of the two TIs become the same with each other at each subsector of 
the blocked Hamiltonian in Eq.(\ref{Qimodel}) because signs of $k_z$  in the two TIs are opposite each other. 

\begin{figure}[htbp]
    \subfigure{\includegraphics*[height=42mm]{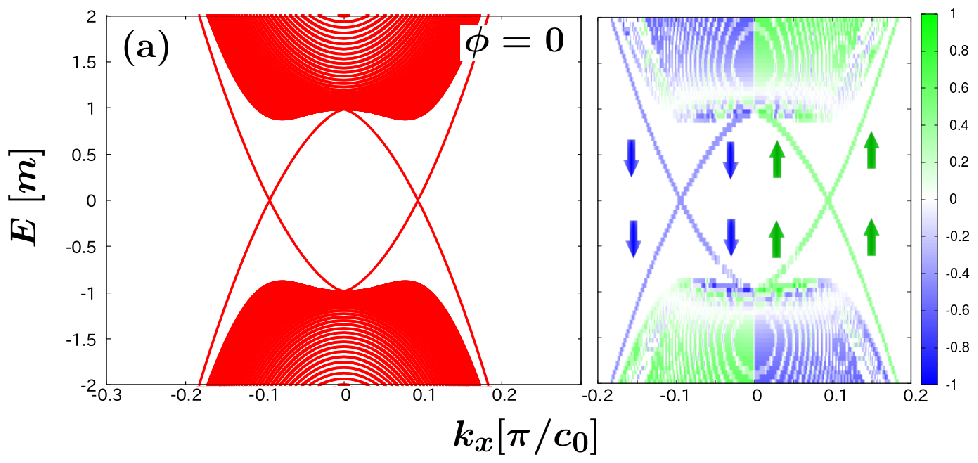}\label{fig:invx-x}}
    \subfigure{\includegraphics*[height=42mm]{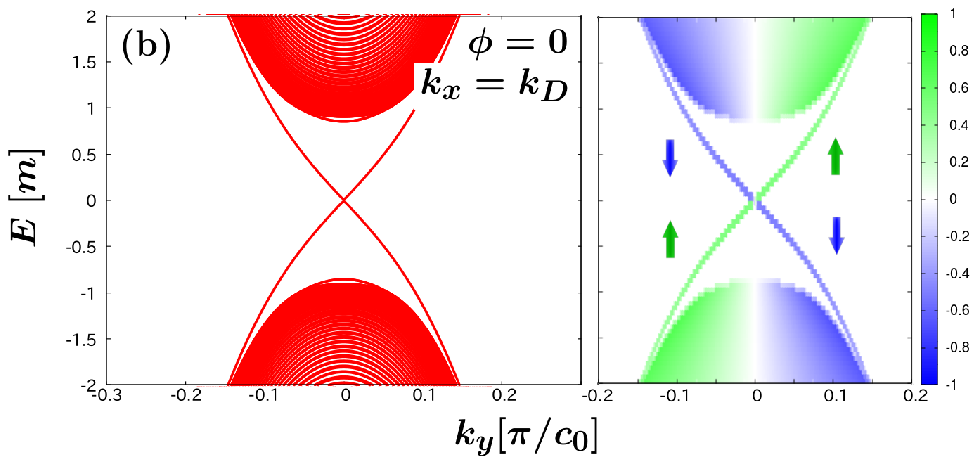}\label{fig:invx-y}}
     \subfigure{\includegraphics*[height=42mm]{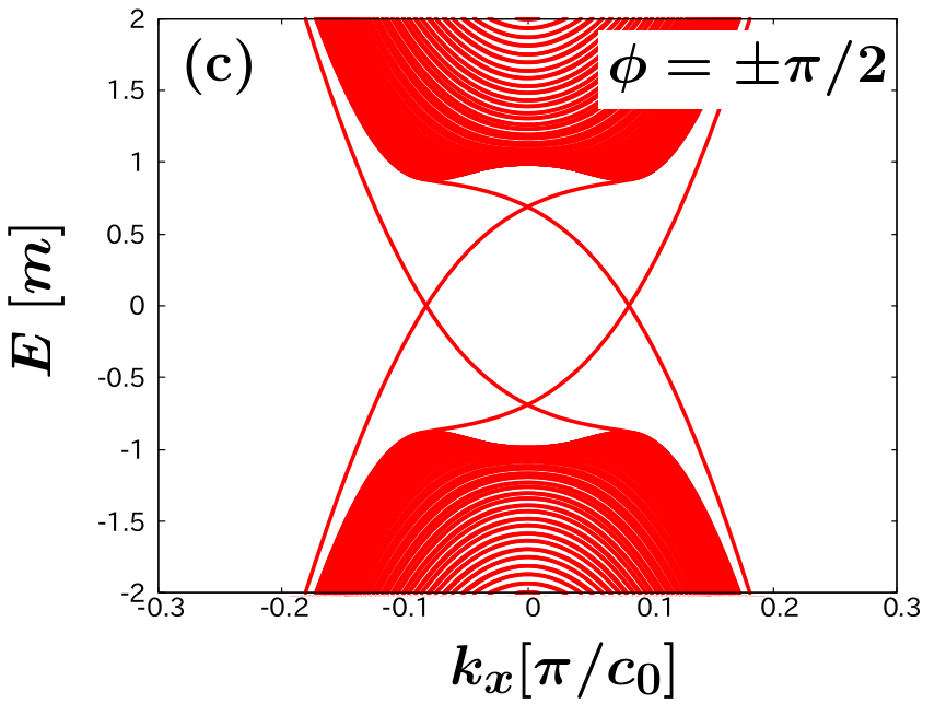}\label{fig:invx-hpi}}
\caption{The energy dispersions of interface states along (a) $k_x$ and (b) $k_y$ are shown at $\phi=0$ and $s=-1$.
In the figure (b), we set $k_x$ at the Dirac point $k_D=0.094\pi$. 
The figure (c) is energy dispersion along $k_x$ at $\phi=\pi/2$ and $s=1$.
 }\label{inv}
\end{figure}

\begin{figure}[htbp]
    \subfigure{\includegraphics*[height=45mm]{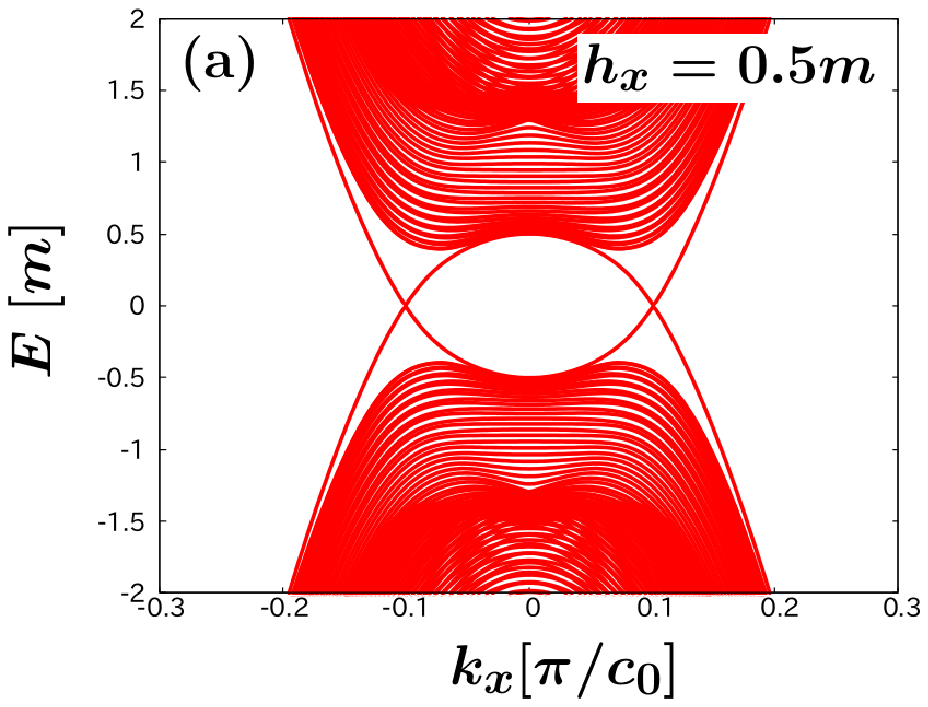}\label{fig:invx+Bx}}
     \subfigure{\includegraphics*[height=45mm]{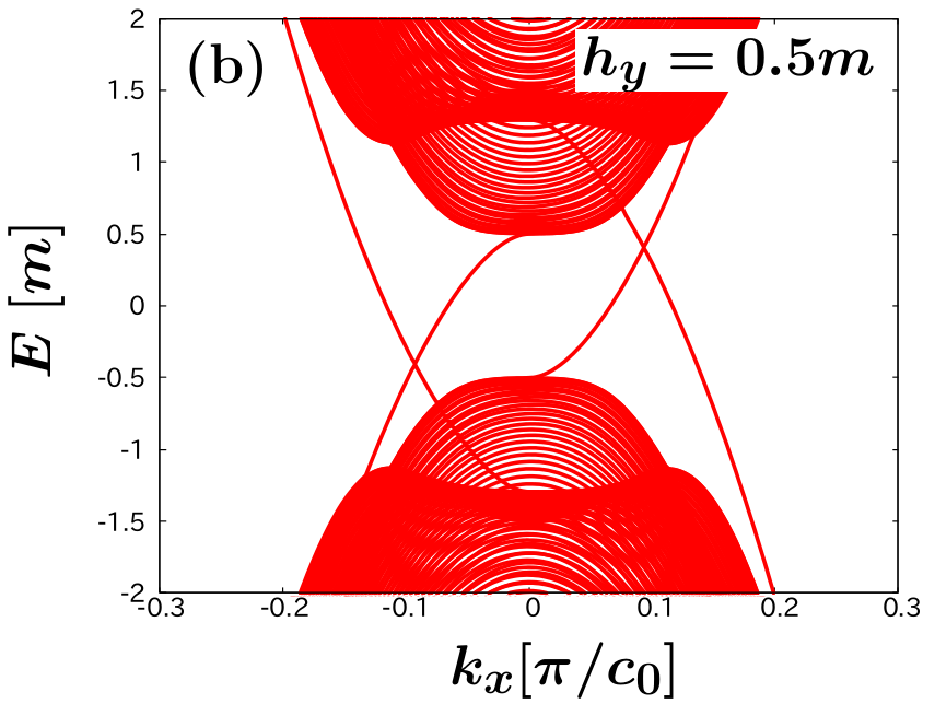}\label{fig:invx+By}}
     \subfigure{\includegraphics*[height=45mm]{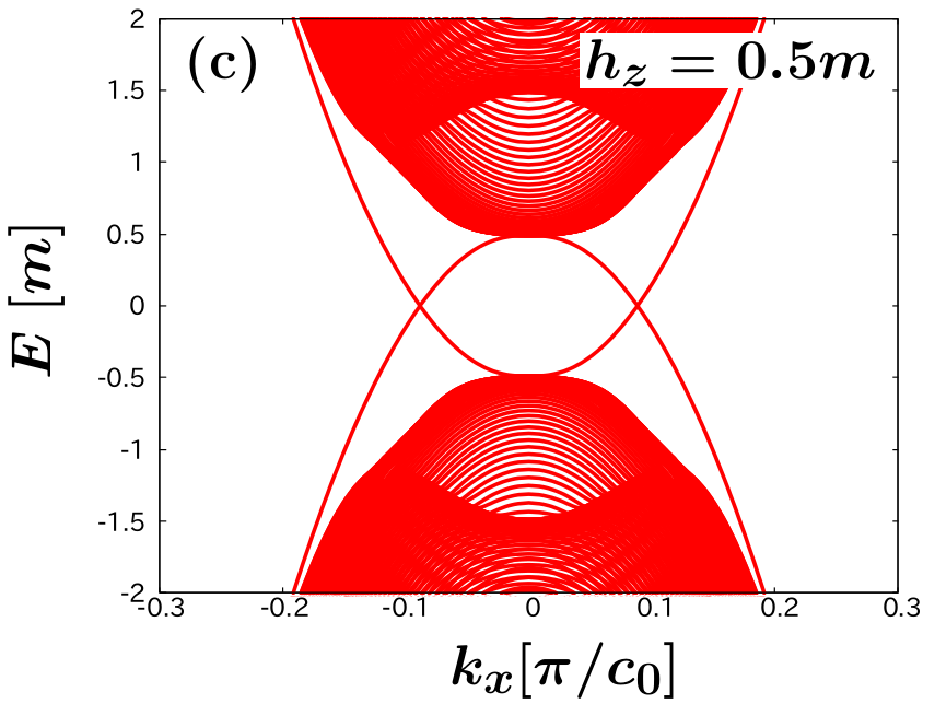}\label{fig:invx+Bz}}
\caption{The energy dispersions of interface states along $k_x$ with the Zeeman field directing $x$-axis (a), $y$-axis (b), and $z$-axis (c) are shown at $\phi=0$ and $s=-1$.
 }\label{fig:10}
\end{figure}

When we choose $s=-1$, Eq.~(\ref{invx}) represents the inversion in the $x$-axis.
The choice of $\phi=\pi$ in Eq.~(\ref{invx}) is identical to Eq.~(\ref{inversion}) with $\phi=\pi$ whose results
are shown in Fig.~\ref{fig:az+pi}. 
Therefore we seek gapless states for $\phi\neq\pi$.
The  gapless states are expected at $k_y=0$ because the relative Chern number well distinguishes the two TIs there.
The results for $\phi=0$ are shown in Figs.~\ref{fig:invx-x} and \ref{fig:invx-y}.
We plot the dispersion as a function of $k_x$ at $k_y=0$ in Figs.~\ref{fig:invx-x}.
The results suggest that there are the two Dirac cones but their nodes stay at finite value of ${k}_x$.
Correspondingly, there is gapless state in the dispersion along the $k_y$ at the Dirac point of $k_x=k_D=0.094\pi$ in Fig.~\ref{fig:invx-y}.
The spin polarization calculated with $S_x$ is also shown in Fig.~\ref{fig:invx-x}.
At $s=-1$, two Dirac cones appear with their nodes staying on $k_x$ axis with all $\phi$ as shown in Fig. \ref{fig:invx-hpi} for $\phi=\pi/2$
and these cones come close to $\boldsymbol{k}=0$ with
 increasing of $\phi$ to $\pi$ in Fig.~\ref{fig:az+pi}.
The analysis of the relative Chern number suggests the robustness against the Zeeman field as discussed in Sec.~\ref{RC} in Fig. \ref{fig:10}. 

For another configurations matrix, the rotation in the $xz$-plane is equivalent to that in the $yz$-plane under interchanging $k_x$ and $k_y$.
Thus the Dirac nodes stay on the $k_y$ axis when we consider the inversion in the $y$ direction.
Together with the results in Fig.~\ref{fig:az+pi}, we conclude that the gapless states appear when the 
relative configuration matrix $\Lambda_{(1)}$ includes the inversion. 
Unfortunately, these gapless states are also fragile under the perturbation 
which breaks the band-inversion symmetry.

\section{conclusion}
We have studied the interface state between two topological insulators with different configurations of spin-orbit 
interactions.
The two topological insulators touch at $z=0$ and its interface is flat within the $xy$ plane.
The coupling of spin $\boldsymbol{\sigma}$ with momenta $\boldsymbol{p}$ 
is configured by a material dependent $3\times 3$ matrix $\boldsymbol{\Lambda}$ as ${\sigma}^\mu {\Lambda}_\mu^\nu p_\nu$.
The spectra of subgap states depends on relative configuration matrix $\boldsymbol{\Lambda}_{12}=\boldsymbol{\Lambda}_{(1)}\boldsymbol{\Lambda}_{(2)}^{-1}$, where $\boldsymbol{\Lambda}_{(l)}$ for $l=1$ and 2 characterize the 
spin-orbit coupling in the two topological insulators.
The configuration matrix $\boldsymbol{\Lambda}_{12}$ consists of two operations: the inversion of some axes 
and the rotation of the angle for the spin-momentum locking.
The gapless interface state appears only when the configuration matrix contains the inversion.
To make clear reasons for the appearing gapless interface states, we introduce two topological numbers defined in the partial Brillouin zone.

The relative Chern number distinguishes the two topological insulators when $\Lambda_{12}$ does not include the inversion along the $z$-axis and explains the appearance of the interface gapless state under the inversion within the $xy$-plane.
On the basis of the numerical calculation, we conclude that such gapless states are robust under the Zeeman field for all spin directions.

Sato's winding number distinguishes the two topological insulators when $\Lambda_{12}$ includes the inversion along the $z$-axis and
explains the robustness of the gapless state under the rotation within the $xy$-plane.
In this case, the four degenerate Dirac points appears at the $\Gamma$ point in the Brillouin zone according to the definition of Sato's winding number.
From the numerical calculation, we conclude the robustness of the Zeeman field within the $xy$-plane. We also conclude that the gapless states are fragile under the Zeeman field in the $z$-direction.

The gapless interface states with the inversions of two axes are fragile under the rotation within the plane including the two axes.
To realize such gapless interface states in experiments, we need the delicate material tuning to fix the angle $\phi$.
Therefore, the gapless state with inversion of a single axis is most robust in the gapless interface states at the junction of two TIs.
Especially, the gapless state with inversion of a single axis parallel to the interface is also robust against the Zeeman field with all directions as shown in Fig. \ref{fig:10}.

We have confirmed all the conclusions of the topological analysis in Sec.~III
and the robustness of the gapless interface state protected topologically by the numerical simulation on the tight-binding model in Sec.~IV.
The numerical simulation shows that all the gapless states are fragile in the presence of the band-inversion 
symmetry breaking perturbations introduced in a different way in the two topological insulators. 
The last property implies the difficulty of finding the gapless states in real materials within the combination 
of existing topological insulators.
However, our results predict an unusual gapless state appearing at the interface of two different
topological superconductors and that of two different superfluid phases.
This is because the Bogoliubov de Gennes Hamiltonian similar to Eq.~(\ref{H2}) always satisfies the 
particle-hole symmetry (band-inversion symmetry in this paper) in Ref.~\onlinecite{Schnyder2008}.\\

{\it Note }. Recently, De Beule and Partoens have studied the interface state in a different point of view.
 They  show the absence of the tachyonic excitation at the interface between two topological insulators.
 The manuscript have been published as the regular paper\cite{Beule2013-2}.

\section{acknowledgments}
The authors are grateful to M.~Sato for useful discussion.
This work was supported by 
the "Topological Quantum Phenomena" (Grant No. 22103002) Grant-in Aid for 
Scientific Research on Innovative Areas from the Ministry of Education, 
Culture, Sports, Science and Technology (MEXT) of Japan.

\appendix

\section{The shift of the Dirac point}\label{AP1}
The effective two-dimensional Hamiltonian of the surface state perpendicular to the $\eta$-axis is represented by
\begin{align}
H_{s}=v_F(\boldsymbol{k}\times\boldsymbol{\sigma}')\cdot \boldsymbol{e}_\eta,
\end{align}
where $v_F$ is the Fermi velocity, $\boldsymbol{k}$ is a two-dimensional momentum, and $\boldsymbol{e}_\eta$ is a unit vector along the $\eta$-axis.
The spin is represented by
\begin{align}
\boldsymbol{\sigma}'=(\Lambda_{\mu}^x,\Lambda_{\mu}^y,\Lambda_{\mu}^z)\sigma^\mu,
\end{align}
where $\Lambda$ is the configuration matrix.
 The index which appears twice in a single term means the summation for $\mu=x,y,z$.
When we consider the surface perpendicular to the $z$-axis under the Zeeman field, for example, the Hamiltonian for the surface states reads
\begin{align}
H=v_F(k_x\Lambda_{\mu}^y-k_y\Lambda_{\mu}^x)\sigma^\mu+B_\mu\sigma^\mu,
\end{align} 
where $B_\mu$ is a magnetic field component for the $\mu$-axis within the plane perpendicular to the $z$-axis.
It is easy to confirm that the Zeeman field shifts the Dirac point from the $\Gamma$ point to the point in the two-dimensional Brillouin zone satisfying
\begin{align}
v_F(k_x\Lambda_{\mu}^y-k_y\Lambda_{\mu}^x)+B_\mu=0.
\end{align}
It is basically possible to observe the shift of the Dirac point by angle resolved photoemission spectroscopy.
Therefore, the configuration matrix $\Lambda_{\mu}^\nu$ can be determined by repeating the measurement for all surfaces.

\section{Mirror operation}\label{APmirror}
We explain the mirror operation about a single plane and two planes normal to each other.
The mirror operation $\mathcal{M}_\mu$ about the plane perpendicular to a single axis of $\mu$ transforms the momentum $p_{\nu}$ and spin $\sigma^{\nu}$ as
\begin{align}
\mathcal{M}_\mu p_{\nu} =&
\begin{cases}
-p_{\nu}&(\nu=\mu)\\
p_{\nu}&(\nu\neq\mu)
\end{cases}\\
\mathcal{M}_\mu \sigma^{\nu}=&
\begin{cases}
\sigma^{\nu}&(\nu=\mu)\\
-\sigma^{\nu}&(\nu\neq\mu)
\end{cases},
\end{align}
where the transformation of the spin $\sigma^{\nu}$ is the same as the angular momentum of $\nu$-axis.
The mirror operation for arbitrary operator $O$ can be represented by an unitary matrix $U$ as $UOU^\dagger$.
The mirror operation $\mathcal{M}_{\mu\nu}$ for two planes perpendicular to the $\mu$ and the $\nu$-axes transforms the momentum $p_{\rho}$ and spin $\sigma^{\rho}$ as
\begin{align}
\mathcal{M}_{\mu\nu} p_{\rho} =&
\begin{cases}
-p_{\rho}&(\rho=\mu\;\mathrm{or}\;\nu)\\
p_{\rho}&(\rho\neq\mu\;\mathrm{and}\;\nu)
\end{cases}\\
\mathcal{M}_{\mu\nu} \sigma^{\rho}=&
\begin{cases}
-\sigma^{\rho}&(\rho=\mu\;\mathrm{or}\;\nu)\\
\sigma^{\rho}&(\mathrm{other\; cases})\\
\end{cases}.
\end{align}
The relations are derived from $\mathcal{M}_{\mu\nu}=\mathcal{M}_{\mu}\mathcal{M}_{\nu}$.
In a junction, the mirror operation is more complicated because it contains exchange the two sides of the junction.
In numerical model, the mirror operations $\mathcal{M}_{\mu}$ and $\mathcal{M}_{\mu\nu}$ exchange the Hamiltonians at the two sides of the junction
when the mirror operations contain that of $z$-axis under the mirror operation $\mathcal{M}_z$. 
\begin{align*}
\mathcal{M}_zH_1(p_\mu,\sigma^\nu)=&H_2(\mathcal{M}_zp_\mu,\mathcal{M}_z\sigma^\nu)\\
\mathcal{M}_zH_2(p_\mu,\sigma^\nu)=&H_1(\mathcal{M}_zp_\mu,\mathcal{M}_z\sigma^\nu).
\end{align*}
The Zeeman field along the $\mu$-axis breaks the mirror symmetry of $\mathcal{M}_\nu$ where the $\nu$-axis is perpendicular to the $\mu$-axis.

\section{Spectra of interfacial bound states}\label{quasi}
We begin with the Hamiltonian in Eq.~(\ref{H2}) which is represented as
\begin{align}
H_0=& \begin{pmatrix}
 M \sigma^0 & a \sigma^\nu\Lambda^\lambda_\nu p_\lambda \\
a\sigma^\nu\Lambda^\lambda_\nu p_\lambda &-M \sigma^0
\end{pmatrix}.\label{h0ti}
%\\
%H_\textrm{{BI}} =& (\mu +c|\boldsymbol{p}|^2) \boldsymbol{1}_{4\times 4}, \label{hbi}\\
%H_{\textrm{TR}}=&
%\left[ \begin{array}{cc} \boldsymbol{h} \cdot \boldsymbol{\sigma} & 0\\
%0 & \boldsymbol{h} \cdot \boldsymbol{\sigma} \end{array} \right],\label{htr}
\end{align}
The Hamiltonian in Eq.~(\ref{h0ti}) preserves 
the band-inversion symmetry.
\begin{align}
\mathcal{D} H(\boldsymbol{p})\mathcal{D}^{-1}=&-H^\ast(-\boldsymbol{p}),\\
\mathcal{D}=&\left(\begin{array}{cc}0 & -\sigma^y \\ \sigma^y & 0 \end{array}\right).
\end{align}
The wave function can be given by
\begin{align}
\Psi^{(+)}_{\boldsymbol{p}}(\boldsymbol{r})
=&\left[ \begin{array}{c} u_p \sigma^0 \\ v_p \Gamma \end{array}\right] e^{i\boldsymbol{p}\cdot\boldsymbol{r}},\,
\Psi^{(-)}_{\boldsymbol{p}}(\boldsymbol{r})
=\left[ \begin{array}{c} -v_p \Gamma \\ u_p \sigma^0  \end{array}\right] e^{i\boldsymbol{p}\cdot\boldsymbol{r}},\\
\Gamma=&\frac{\sigma^\nu\Lambda^\lambda_\nu p_\lambda}{|\boldsymbol{p}|},\quad E_p=\sqrt{M^2+a^2|\boldsymbol{p}|^2}\\
u_p=&\sqrt{\frac{1}{2}\left(1+\frac{M}{E_p}\right)}, \quad
v_p=\sqrt{\frac{1}{2}\left(1-\frac{M}{E_p}\right)},
\end{align}
where $\Psi^{(+)}$ and $\Psi^{(-)}$ are the wave functions belonging to $E_p$ and $-E_p$, respectively.
It is evident that $E_p$ is independent of ${\Lambda}^\mu_\nu$.

When two topological insulators touch each other at $z=0$, 
the wave function for $E=E_p>0$ is represented by
\begin{align}
\Psi^{(1)}(\boldsymbol{r})
= &\left[ \left(\begin{array}{c} v \sigma^0 \\ u \Gamma_{(1)}^- \end{array}\right) e^{-ik^+_zz} \hat{A}
+\left( \begin{array}{c} u \sigma^0 \\ v \Gamma_{(1)}^+ \end{array}\right) e^{ik^-_zz} \hat{B}
\right]\nonumber\\
&\times e^{i\boldsymbol{k}\cdot\boldsymbol{\rho}},
\end{align}
for $z<0$ and
\begin{align}
\Psi^{(2)}(\boldsymbol{r})
= &\left[ \left(\begin{array}{c} u \sigma^0 \\ v \Gamma_{(2)}^- \end{array}\right) e^{-ik^-_zz} \hat{C}
+\left( \begin{array}{c} v \sigma^0 \\ u \Gamma_{(2)}^+ \end{array}\right) e^{ik^+_zz} \hat{D}
\right] \nonumber\\
&\times e^{i\boldsymbol{k}\cdot\boldsymbol{\rho}},
\end{align}
for $z>0$ with
\begin{align}
&\Gamma^\pm_{(l)} = \frac{1}{|\boldsymbol{p}|} \sigma^\nu (\Lambda_{(l)})_\nu^\lambda (p_\pm)_\lambda,
\quad \boldsymbol{p}_\pm=(\boldsymbol{k},\pm k_z),\\
%(\Lambda^1_\nu k_x + \Lambda^2_\nu k_y \pm \Lambda^3_\nu k_z),\\
u=  &\sqrt{\frac{1}{2}\left(1+\frac{\Omega}{E}\right)}, \quad
v=  \sqrt{\frac{1}{2}\left(1-\frac{\Omega}{E}\right)},\\
\Omega=&\sqrt{E^2-E_g^2}, \quad k_z^{\pm}=k_0\sqrt{ 1\pm \frac{\Omega}{bk_0^2}},
\end{align}
\begin{align}
\boldsymbol{p}=&(\boldsymbol{k},k_z), \quad \boldsymbol{k}=(k_x,k_y),\quad \boldsymbol{\rho}=(x,y),\\
p_F=&\sqrt{\frac{m}{b}-\frac{a^2}{2b^2}}, \quad
k_0= \sqrt{ p_F^2-|\boldsymbol{k}|^2 }, \\
 E_g=&a\sqrt{\frac{m}{b}-\frac{a^2}{4b^2}},
\end{align}
where $l=1$ and 2 indicate the two topological insulators.
The $2\times 1$ coefficient matrices $\hat{A}$, $\hat{B}$, $\hat{C}$, and $\hat{D}$ represent the amplitude of outgoing 
waves from the interface.
At a surface of a topological insulator, it is easy to confirm that 
the spectra of the surface state become
$E= a |\boldsymbol{k}|$ from the boundary condition $\left.\Psi^{(1)}(\boldsymbol{r})\right|_{z=0}=0$. 

At the interface of the two topological insulators,
the boundary conditions at $z=0$, 
\begin{align}
\Psi^{(1)}(\boldsymbol{\rho},0)=&\Psi^{(2)}(\boldsymbol{\rho},0), \\
\left. \partial_z\Psi^{(1)}(\boldsymbol{\rho},z)\right|_{z=0}=&\left. 
\partial_z \Psi^{(2)}(\boldsymbol{\rho},z)\right|_{z=0},
\end{align}
gives a condition
\begin{align}
&\mathrm{det}\left[
(k_z^+-k_z^-)^2u^2v^2AB-4k_z^+k_z^-ACA^{-1}D\right]=0,\label{c1}
\end{align}
where $A$, $B$, $C$, and $D$ are represented by
\begin{align*}
A=&\Gamma_{(2)}^+-\Gamma_{(1)}^-\\
B=&\Gamma_{(2)}^--\Gamma_{(1)}^+\\
C=&u^2\Gamma^+_{(2)}-v^2\Gamma^+_{(1)}\\
D=&u^2\Gamma^-_{(1)}-v^2\Gamma^-_{(2)}.
\end{align*}
In the quasiclassical approximation, 
we use a relation $a^2\ll mb$, which allows 
$|k_z^+-k_z^-|^2\ll k_z^+k_z^- \approx k_0^2$.
Thus we find 
\begin{align}
\mathrm{det}&\left[ u^2\Gamma^-_{(1)}-v^2\Gamma^-_{(2)}\right]
%\nonumber \\&\times
\mathrm{det}\left[ u^2\Gamma^+_{(2)}-v^2\Gamma^+_{(1)}\right]=0. \label{c2}
\end{align}
From the condition, we obtain the energy of interface subgap states
\begin{align}
E= a p_F \sqrt{ \frac{ |\boldsymbol{p}|^2 + (p_\pm)^\nu \left[{\Lambda}_{(1)}
{\Lambda_{(2)}}^{-1}\right]_\nu^\lambda (p_\pm)_\lambda}{2|\boldsymbol{p}|^2}}. \label{spectra}
\end{align} 
The results indicate the spectra of subgap interface states depend on relative choice 
of ${\Lambda}_{(1)}$ and ${\Lambda}_{(2)}$. 
Within Eq.~(\ref{h0ti}), the spectra obtained from Eq.~(\ref{c2}) shown in Eq.~(\ref{spectra}) 
always doubly degenerate.

When we fix We fix $\Lambda_{(2)}$ at $\textrm{diag}(1,1,1)$ and $\Lambda_{(1)}$ in Eq.~(\ref{mirror}), 
we obtain 
\begin{align}
E= \pm a\sqrt{{p_F}^2-|\boldsymbol{k}|\sin^2(\phi/2)}.
\end{align}
When we choose $\Lambda_{(1)}$ as shown in Eq.~(\ref{inversion}), we find
\begin{align}
E=\pm a|\boldsymbol{k}||\cos(\phi/2)|, 
\end{align}
for $\phi\neq \pi$. The expression is not valid at $\phi=\pi$ because Eq.~(\ref{c2}) 
is always satisfied due to the quasiclassical approximation. Thus we need to solve 
Eq.~(\ref{c1}) to obtain the spectra. By considering the first order correction 
proportional to $a^2/mb$, we obtain the spectra of the linear dispersion at $\phi=\pi$
\begin{align}
E=\pm \sqrt{\frac{a^2}{4mb}} a |\boldsymbol{k}|.
\end{align}
To obtain the spectra with the cubic dispersion, we need to consider higher order corrections.

\section{Effects of disorder}\label{AP4}
 We have introduced the topological numbers to distinguish two topological insulators. 
These topological numbers are defined in the partial Brillouin zone specified by a wavenumber. 
In the presence of disordered potential, such partial Brillouin zones are not well defined. Therefore 
the topological distinction may not work well. 
To confirm the instability of the interface states under the disordered potential, we numerically calculate 
the energy spectra of the Hamiltonian in the presence of disorder. Here we show a result of numerical calculation.
\begin{figure}[htbp]
  \subfigure{\includegraphics*[width=70mm]{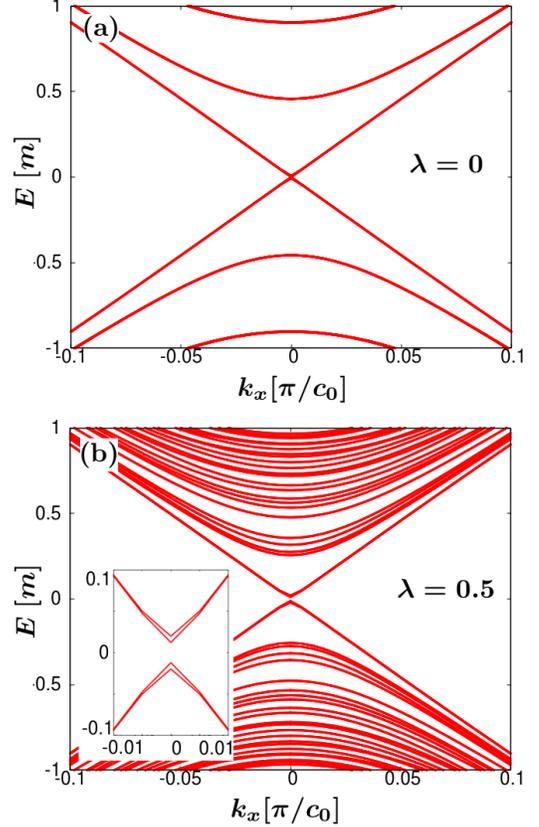}}
\caption{We compare the energy spectra of the interface states in the absence of disorder (a) with those in the presence of disorder (b).
Strictly speaking, there is a very small gap in the subgap spectra even in (a) due to to the finite size effect. 
However, in (b), the gap in the presence of the disorders becomes larger than that in the absence of the disorder}\label{figap1}
\end{figure}
To avoid the finite size effect, we introduce the quasi-random potential which is random within the $yz$ plane but uniform
in the $x$ axis. In this model, $k_x$ is still a good quantum number but $k_y$ and $k_z$ are no longer good 
quantum numbers.
We consider the tight-binding model in the two-dimensional real space in the $yz$ plane $\boldsymbol{j}=(j_1,j_2)$ with the 
one-dimensional momentum $k_x$
\begin{align}
&H_{0}(k_x;(j_1,j_2),(j'_1,j'_2)) = 
\begin{pmatrix}
M{\sigma}^0 & h_{\textrm{so}}\\
h_{\textrm{so}} & -M{\sigma}^0
\end{pmatrix},\label{tbh0}\\
&M = (m +2b_2 (\cos(k_xc_0)-2))\delta_{\boldsymbol{j},\boldsymbol{j}'}\nonumber\\
&\qquad- 2b_1\delta_{\boldsymbol{j},\boldsymbol{j}'}+ b_1 (\delta_{\boldsymbol{j},\boldsymbol{j}'+\boldsymbol{e}_z} + \delta_{\boldsymbol{j},\boldsymbol{j}'-\boldsymbol{e}_z})\nonumber\\
&\qquad- 2b_2 (r_{\boldsymbol{j}}^+\delta_{\boldsymbol{j},\boldsymbol{j}'+\boldsymbol{e}_y} + r_{\boldsymbol{j}}^-\delta_{\boldsymbol{j},\boldsymbol{j}'-\boldsymbol{e}_y})
, \\
&h_{\textrm{so}}=
a_2 \sigma^\nu\left[ \Lambda_\nu^x\sin(k_xc_0) +\Lambda_\nu^y\sin(k_yc_0)\right]\delta_{\boldsymbol{j},\boldsymbol{j}'} \nonumber\\
&\qquad-ia_2\sigma^\nu\Lambda_\nu^y(\delta_{\boldsymbol{j},\boldsymbol{j}'+\boldsymbol{e}_y}-\delta_{\boldsymbol{j},\boldsymbol{j}'-\boldsymbol{e}_y})\nonumber\\
&\qquad-ia_1 \sigma^\nu \Lambda_\nu^z(\delta_{\boldsymbol{j},\boldsymbol{j}'+\boldsymbol{e}_z}-\delta_{\boldsymbol{j},\boldsymbol{j}'-\boldsymbol{e}_z}),
\end{align}
with
\begin{align}
r_{\boldsymbol{j}}^+=r_{\boldsymbol{j}+\boldsymbol{e}_y}^-=1+\epsilon_{\boldsymbol{j}},
\end{align}
where $\boldsymbol{e}_\mu$ is the unit vector along the $\mu$ axis.
The perturbation on the hopping $\epsilon_{\boldsymbol{j}}$ is given randomly in the range of $[-\lambda,\lambda]$ at each lattice site $\boldsymbol{j}$.
The lattice sizes are $N_y=40$ and $N_z=20$, where $N_\mu$ is the number of lattice points in the $\mu$ axis.
We employ the periodic boundary condition along the $y$ and $z$ axes.
We consider $\Lambda_{(1)}$ in Eq.~(\ref{inversion}) with $\phi=0$ in Sec. III B, where the definition of Sato's number requires 
the partial Brillouin zone specified by $(k_x, k_y)=(0,0)$.
The energy spectra of interface states for $\lambda=0$ and $\lambda=0.5$ are shown in Fig\ref{figap1}(a) and (b) respectively.
It is clear that the disordered potential induces the gap in the spectra of the interface states. 
We have also confirm that the gap also appears in the interface energy spectra 
for the case of topological states with the relative Chern number in Sec. III A.
Therefore the topological distinction in terms of the topological number defined in the partial Brillouin zone 
does not work well in the presence of disorder.

\bibliography{TI}

\end{document}